\documentclass[prc,preprint,amsmath,amssymb,
floatfix]{revtex4-1}
\usepackage[utf8]{inputenc}
\usepackage{amsfonts}
\usepackage{graphicx}
\usepackage{fullpage}
\usepackage{mathtools}
\usepackage{bm}
\usepackage{subfig}

\usepackage{CJKutf8}
\usepackage{mathrsfs}
\usepackage{bbm}
\usepackage{graphicx}
\usepackage{dcolumn}
\usepackage{bm}
\usepackage{float}
\usepackage{color}
\usepackage{multirow}
\usepackage{enumerate}
\usepackage{BOONDOX-cal}
\usepackage{listings}
\usepackage{xcolor}

\begin{document}
%\begin{CJK*}{GB}{} % Use default fonts from CJK (see below)

\title{Nuclear states projected from a pair condensate }

\author{Yi Lu }
%\author{Yi Lu (路毅) }
\affiliation{College of Physics and Engineering, Qufu Normal University, Jingxuan West Road 57, Qufu, Shandong 273165, China}

%\author{Yang Lei (雷杨) }
\author{Yang Lei }
\email[corresponding author: ]{leiyang19850228@gmail.com}
\affiliation{School of National Defense Science and Technology, Southwest University of Science and Technology, Mianyang 621010, China}

\author{Calvin W. Johnson }
\email[corresponding author: ]{cjohnson@sdsu.edu}
\affiliation{ Department of Physics, San Diego State University, 5500 Campanile Drive, San Diego, CA 92182, USA}

%\author{Jia Jie Shen (沈佳杰)}
\author{Jia Jie Shen}
\affiliation{ School of Arts and Sciences, Shanghai Maritime University, Shanghai 201306, China }

\begin{abstract}
Atomic nuclei exhibit deformation, pairing correlations, and rotational symmetries. To meet these competing demands in a computationally tractable formalism,  
we revisit the use of general pair condensates with good particle number as a trial wave function for even-even nuclei. After minimizing the energy of the condensate, allowing for general triaxial deformations, 
 we project out states with good angular momentum with a fast projection technique. 
To show applicability, we present example calculations from pair condensates in several model spaces, 
and compare against  angular-momentum projected Hartree-Fock and full configuration-interaction shell model calculations.
This approach successfully generates  spherical, vibrational and rotational spectra, demonstrating potential for modeling medium- to heavy-mass nuclei.
\end{abstract}

\maketitle

\section{Introduction}\label{sec-int}

Medium- and heavy-mass nuclides exhibit a rich portfolio of behaviors, not only the classic rotational, vibrational, and superfluid (spherical) spectra~\cite{bohr1998nuclear,Ring1980}, but also 
 triaxial deformation and wobbling~\cite{PhysRevC.89.014322} and chiral doublet bands~\cite{PhysRevLett.110.172504}.   An accurate description of the many-body wave functions
 of heavy nuclides is important to understanding  alpha cluster pre-formation in $\alpha$-decay~ \cite{PhysRevC.95.061306} as well as nucleosynthesis in astrophysical environments~\cite{Surman2016The,KAJINO2019109}.

There are numerous microscopic approaches to modeling heavy nuclei, including  the relativistic mean field method~\cite{2016Relativistic}, the Hartree-Fock Bogoliubov method~\cite{Ring1980,PhysRevC.103.024315,PhysRevC.104.054306,PhysRevC.100.044308,PhysRevC.103.064302},  
the random phase and related approximations~\cite{Ring1980}, and  algebraic models  (e.g. the Interacting Boson Model \cite{Iachello2006}), as well as the projected shell model \cite{PhysRevC.61.064323} and  powerful Monte Carlo methods~\cite{mcsm-pair}.  

Here we focus on the configuration-interacting shell model, which  has been successful in reproducing low-lying states and transition rates of light and medium-heavy nuclei \cite{RevModPhys.77.427, BROWN2001517}.
Within a given valence space and 
with a good effective interactions, the configuration-interaction  shell model reproduces large amounts of data.
The downside of configuration-interaction methods, however, is they are at the mercy of the valence space.  While any calculation in the $1s0d$ shell is nowadays nearly trivial~ \cite{PhysRevC.74.034315} 
and much of the $1p0f$ shell is accessible ~\cite{PhysRevC.65.061301,Honma2005}, as one reaches up to the $2s1d0g_{7/2}0h_{11/2}$ space (nuclides with $N,Z \in [50, 82]$),
calculations quickly become intractable due to  the large dimension of the many-body basis.

Among many different configuration-interaction truncation schemes, one is to construct the basis from products of a restricted set of fermion pairs coupled to angular momentum $0, 2, 4, 6, \cdots$ (S, D, G, I pairs).
In low-lying states, such pairs are  favored energetically.
An example of this is the general seniority model, which focuses on S-pair condensates and their breaking \cite{1981Simple,PhysRevC.99.014302,PhysRevC.93.064307,Jia_2015}, and is especially useful in semi-magic nuclei.
Going further, the nucleon-pair approximation (NPA) truncates the configuration space with angular-momentum coupled fermion pairs, and has proved to be effective in medium-heavy and heavy nuclei with schematic pairing-plus-quadrupole-quadrupole interactions  \cite{ZHAO20141, PhysRevC.65.054317}.   In the NPA matrix elements and overlaps between many-pair states with good total angular 
momentum (i.e., $J$-scheme) are computed using commutation 
relations, which become numerically burdensome as the number of pairs increase, in practice limiting calculations to a maximum of about 8 valence pairs.  To address this burden,
 $M$-scheme NPA codes have been developed recently, pushing the upper limit of valence particles in the NPA  to  $Z_\mathrm{valence}, N_\mathrm{valence} \leq 12$ \cite{PhysRevC.102.024304,Lei2021}.

An alternate is to replace fermion pairs by bosons.  
The Interacting Boson Model (IBM)~\cite{Iachello2006} typically starts from  S- and D-bosons and constructs Lie group reduction chains, classifying nuclei into different group limits.
There have been  efforts to map shell model calculations to boson models or pair models ~\cite{OTSUKA1978139,PhysRevC.50.R571,Ginocchio19951861}. 
While bosons models have the advantage of simpler commutation relations than fermion pairs, rigorous mapping of fermion pairs into bosons is far from trivial. 

Related to the challenge of mapping fermion pairs to bosons is the problem of choosing the structure of the fermion pairs. 
 Pairs determined from the variational principle, either from varying pairs by direct conjugate gradient descent, or extracted from a deformed Hartree-Fock (i.e., single-particle variation)  state,   can  produce 
 good quality spectra and transitions of transitional and deformed nuclei~\cite{FU2020135705,PhysRevC.104.024312,PhysRevC.102.024310}. 
In other words, it can be useful to combine a variational approach (for optimization) and pair truncations (for small dimension).

Variational methods  start from a given trial wave function  (or ansatz) and minimizes the expectation value of the given Hamiltonian.
Hartree-Fock calculations use a  Slater determinant, an antisymmetrized product of single-particle states, as the ansatz, and varies the single-particle states to minimize the energy.
The Hartree-Fock-Bogoliubov scheme utilizes a quasi-particle (Bardeen-Cooper-Schrieffer or BCS) vacuum constructed from a canonical basis as the trial wave function, and the canonical basis as well as BCS occupancies are varied~\cite{Ring1980}.
 Slater determinants used in Hartree-Fock calculations have the advantage of  fast evaluation of Hamiltonian matrix elements, while the HFB scheme benefits from inherent simplicity of the quasi-particle vacuum.

Exploration of ans\"atze  is important and can lead to new approaches~\cite{PhysRevC.104.L031305}.
In this paper we focus on a pair condensate ansatz,  well known in quantum chemistry as ``antisymmetrized geminal power'' wave functions~\cite{Coleman1965},
\begin{equation}
(\hat{A}^\dagger)^n | 0 \rangle,~~~~ \hat{A}^{\dagger}=\frac{1}{2}\sum_{\alpha \beta} A_{\alpha \beta} \hat{c}^{\dagger}_\alpha \hat{c}^{\dagger}_\beta,
\label{eqn-pair-condensate}
\end{equation}
where $|0\rangle$ is the bare fermion vacuum, and $\hat{c}^{\dagger}_\alpha, \hat{c}^{\dagger}_\beta$ are fermion creation operators with $\alpha, \beta$ labelling single-particle states  (i.e. $\alpha$ stands for $(n_\alpha l_\alpha j_\alpha m_\alpha)$), and, finally, $A_{\alpha \beta}$ are the ``pair structure coefficients."
%In this work we denote all operators with hatted symbols, like $\hat{A}^\dagger$, while corresponding structure coefficient matrices are denoted with plain capital letters, like $A$.
  This is also called a ``collective'' pair, although sometimes that term is restricted 
to pairs with good angular momentum; we make no such restriction here.  When we find a variational minimum, 
we restrict ourselves to real pair structure coefficients with skew symmetry $A_{\alpha \beta} = - A_{\beta \alpha}$; however later under rotation we will have complex skew-symmetric structure coefficients. 
With proper choice of $A_{\alpha\beta}$, such a condensate wave can be a Slater determinant as in Hartree-Fock theory \cite{Ring1980,FU2020135705}, or a seniority-0 state in Talmi's generalized seniority model~\cite{1981Simple}, or particle-number-projected component of a BCS or HFB vacuum with particle number $2n$~\cite{PhysRevC.44.R598}. 
Therefore the pair condensate is a more general ansatz than any one of those limits.

The pair condensate ansatz has not been widely used in nuclear structure theory, however, because the evaluation of Hamiltonian matrix elements to be varied can be complicated \cite{PhysRevC.44.R598, CHEN1995181, PhysRevC.102.024310}, and the angular-momentum projection from a pair condensate has not been previously implemented for nuclear structure.
In recent work~\cite{Lei2021}, we rederived formulas (see \cite{PhysRevC.50.R571,Ginocchio19951861} and references therein) for general $M$-scheme nucleon-pair bases,
\begin{equation}
 \hat{A}^\dagger_1 \cdots \hat{A}^\dagger_n | 0 \rangle,
\end{equation}
with the pair condensate (\ref{eqn-pair-condensate}) as  a special case.
%Note that in Ref. \cite{Lei2021} we used the notation $\hat{P}^\dagger$ for a pair, while in this work we use $\hat{A}^\dagger$ to differentiate it from projection operator notation $\hat{P}^J_{KM}$.
 In this work we  present the explicit formulas for pair condensates, and implement angular momentum projection after variation.
Variation after angular momentum projection (VAP) leads to lower energies but are more complicated and time-consuming; hence we restrict ourselves to projection after variation (PAV).
Hereafter, we label this approach as ``projection after variation of a pair condensate" (PVPC).
It is  straightforward to introduce more pair condensates into the configuration space, such as in the generator coordinate method~\cite{Ring1980}, but for now we only consider one pair condensate.

In Sec. \ref{sec-formalism}, we outline the formalism for variation of a pair condensate, with details in the appendix. We show typical timing to find the minimum, as variation is the most time consuming part of the calculation, as exemplified by the descent of $^{132}$Dy in the $2s1d0g_{7/2}0h_{11/2}$ shell.
In Sec. \ref{subsec-projection}, we briefly introduce the formalism for angular momentum projection, following a recently developed recipe: linear algebraic projection (LAP) \cite{PhysRevC.96.064304,2019Convergence}.
In Sec. \ref{subsec-decomposition}, we decompose a pair condensate into fractions with good angular momentum, and show typical probability distributions for rotational nuclei and spherical nuclei.
In Sec. \ref{subsec-EMtrans}, we outline the fomalism for electromagnetic transitions between projected wavefunctions.
In Sec. \ref{sec-results}, we present benchmark calculations of nuclides in the $1s0d, 1p0f$ and $2s1d0g_{7/2}0h_{11/2}$ valence spaces with discussions and analysis.
Finally we summarize and discuss future directions in Sec. \ref{sec-sum}.

\section{Formalism} 
\label{sec-formalism}

For the nuclear many-body Hamiltonian, we use shell model effective interactions in occupation space, with 1- and 2-body parts, in proton-neutron format,
\begin{equation}
\begin{aligned}
\hat{H} = & \hat{H}_0 + \hat{H}_{pp} + \hat{H}_{nn} + \hat{H}_{pn} \\
    & = \sum_{a \in \pi} \epsilon_a \hat{n}_a + \sum_{a \in \nu} \epsilon_a \hat{n}_a    \\
    & + \sum_{abcd \in \pi} \frac{\sqrt{(1+\delta_{ab})(1+\delta_{cd})}}{4} 
 \sum_{I} V_{pp}(abcd;I) \sum_{M} \hat{A}^\dagger_{I M}(ab) \hat{A}_{IM}(cd)    \\
    & + \sum_{abcd \in \nu} \frac{\sqrt{(1+\delta_{ab})(1+\delta_{cd})}}{4} 
\sum_{I} V_{nn}(abcd;I) \sum_{M} \hat{A}^\dagger_{I M}(ab) \hat{A}_{IM}(cd)           \\    
    & + \sum_{ac\in \pi, bd \in \nu,I} V_{pn}(abcd;I)
\sum_{M} \hat{A}^{\dagger}_{IM}(a b)
\hat{A}_{IM}(cd).
\end{aligned}
\label{eqn-hamiltonian}
\end{equation}
where $\hat{A}^\dagger_{I M}(ab)$ and  $\hat{A}_{IM}(cd)$ are so-called ``non-collective" pair creations and annihilation operators (in contrast to 
the collective pair creation operator of Eq.~(\ref{eqn-pair-condensate})): 
$\hat{A}^\dagger_{IM}(ab) = \left [\hat{c}^\dagger_a \otimes \hat{c}^\dagger_b \right ]_{IM},$ where $\otimes$ signals coupling via Clebsch-Gordan coefficients, 
and $\hat{A}_{IM}(ab) = \left ( \hat{A}^\dagger_{IM}(ab) \right)^\dagger$. 

We work in several model spaces: in the $1s_{1/2}$-$0d_{3/2}$-$0d_{5/2}$ or $1s0d$ shell, using the universal $1s0d$-shell interaction version B, or USDB ~ \cite{PhysRevC.74.034315}; 
in the $1p_{1/2}$-$1p_{3/2}$-$0f_{5/2}$-$0f_{7/2}$ or $1p0f$ shell, using a monopole-modified $G$-matrix interaction version 1A, or GX1A~\cite{PhysRevC.65.061301, Honma2005}; and finally,
for the space between magic numbers 50 and 82, comprising the orbits $2s_{1/2}$-$1d_{3/2}$-$1d_{5/2}$-$0g_{7/2}$-$0h_{11/2}$, or $2s1d0g_{7/2}0h_{11/2}$, 
 with a monopole-modified $G$-matrix interaction~\cite{PhysRevC.86.044323}.

In order to carry out our calculations, we need a variety of matrix elements. To begin with, for variation (before projection) we need for a given state $| \Phi \rangle$: 
the normalization $\langle \Phi | \Phi \rangle$; the expectation value of a Hamiltonian or the energy of that state $ E(\Phi) = \langle  \Phi | \hat{H} | \Phi \rangle / \langle \Phi | \Phi \rangle$; and the variation 
of the energy with respect to some parameter $\partial E( \Phi ) /\partial \lambda$. To carry out projection, we need the overlap or norm kernel of the state $\Phi$ with a rotated state $\Phi^\prime$, 
$\langle \Phi | \Phi^\prime \rangle$, the Hamiltonian matrix element or Hamiltonian kernel between these two states, $\langle \Phi | \hat{H} | \Phi^\prime \rangle$, and, finally, in order to 
calculate electromagnetic and other transitions represented by a one-body operators $\hat{Q}$, we need $\langle \Phi | \hat{Q} | \Phi^\prime \rangle$.  

All of these are easily found for, say, Hartree-Fock and Hartree-Fock-Bogoliubov calculations.  For the pair condensate calculations the formalism is more complicated and less intuitive. We present 
the detailed formalism in the Appendix, but sketch out the application here.

Restricting ourselves to even-even nuclei, our trial ansatz is a direct product of proton and neutron pair condensates
\begin{equation}
| \Phi \rangle =  (\hat{A}^\dagger_{\pi} )^{N_\pi/2} (\hat{A}^\dagger_{\nu})^{N_\nu/2} | 0 \rangle, \label{eqn:pc-configuration}
\end{equation}
where $N_\pi, N_\nu$ are valence numbers for protons and neutrons.
We vary $|\Phi\rangle$, or more exactly, vary the proton and neutron pair structure coefficients, to minimize the energy
\begin{equation}
\frac{ \langle \Phi | \hat{H} | \Phi \rangle }{ \langle \Phi | \Phi \rangle }.
\end{equation}

In order to find minima, 
we apply  GNU scientific library (GSL) \cite{gsl} conjugate gradient tools.
We do variation before projection, that is, we first find the minimum energy of the unprojected pair condensate, and afterwards project out states of good angular momentum. 

\begin{figure}[h!]
\includegraphics[width=0.7\linewidth]{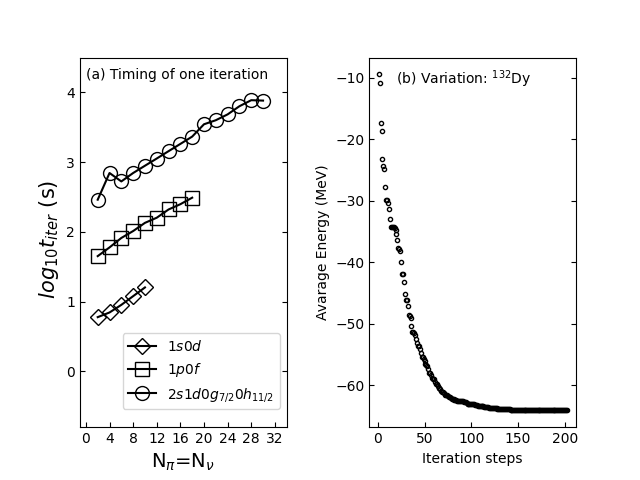}
\caption{
Panel (a): Logarithm CPU time (s) of one iteration in the variation for $N=Z$ nuclei, using conjugate gradient method\cite{1984Practical} wrapped in the GNU Scientific Library function $\it gsl\_multimin\_fdf\_minimizer$\cite{gsl}. $1s0d$, $1p0f$, $2s1d0g_{7/2}0h_{11/2}$ are three valence shells. All these data are collected on a 44-core small workstation, in parallel with simple OpenMP.
Panel (b): The descent of the average energy $\langle \Phi | \hat{H} | \Phi \rangle / \langle \Phi | \Phi \rangle$ for $^{132}$Dy as a test case in $2s1d0g_{7/2}0h{11/2}$ shell. There are $32(32-1) = 992$ free parameters for the variation of the proton and neutron pairs. }
\label{fig:timer}
\end{figure}

Fig. \ref{fig:timer}(a) shows the central processing unit (CPU) time for one iteration for $N=Z$ nuclei in different major shells, with increasing valence particles. 
Our timing data were taken on a $44$-core  workstation using OpenMP parallelization.
In the $1s0d, 1p0f, 2s1d0g_{7/2}0h_{11/2}$ major shells, the numbers of free parameters in variation of one pair condensate basis are $132, 380,$ and $992$, respectively.
The computation complexity of variation has polynomial dependence on the valence particle numbers.
In Fig. \ref{fig:timer} (b), an example is given for variation of $^{132}\mathrm{Dy}$ in the $2s1d0g_{7/2}0h_{11/2}$ shell, with interactions based on the CD-Bonn potential with the G matrix approach and monopole optimizations \cite{PhysRevC.86.044323}.
With 992 free parameters, the iteration tends to converge after 200 steps.
We note that the descent is steep for the first 50 steps, and less so afterwards.
The descent terminates when the modulus of the gradient vector is smaller than a given criterion.
If we set a stricter criterion on the gradient, the descent will continue beyond 200 steps, but with only a slight additional lowering of the energy.
Starting from several independent random pairs, we typically find 2 or 3 local minima.

\subsection{Angular momentum projection} \label{subsec-projection}

In this work we allow for triaxial deformations. 
While there is extensive literature on the standard quadrature approach for angular momentum projection \cite{Ring1980, Sun2016, J.M.Yao2010, R.Rodriguez-Guzman2002}, we adopt the
recently developed  ``linear algebraic projection"(LAP) technique ~\cite{PhysRevC.96.064304,2019Convergence}.
Although both approaches are  based on standard quantum theory of angular momentum \cite{Edmonds1957}, i.e. rotational properties of irreducible tensors, for our application we found  LAP to be 10 times faster than quadrature.

For the standard development of angular momentum projection see \cite{Edmonds1957,Ring1980}.  The basic idea is that a pair condensate state, or any state 
$|\Phi \rangle$  for that matter,  can be decomposed with normalized spherical tensor components, with total angular momentum $I$ and angular momentum projection on the 3rd axis $K$,
\begin{equation}
| \Phi \rangle = \sum_{IK} c_{IK} | IK \rangle. \label{eqn-tensor-decomposition}
\end{equation}
In any finite space the sums are not infinite but restricted to a maximum $I$, $K$, and in practice the coefficients $c_{IK}$ restricts the sum even further~\cite{PhysRevC.96.064304}.
The projection operator $\hat{P}^J_{MK}$ picks out the $(J,K)$ spherical tensor component and rotates it to a $(J,M)$ spherical tensor,
\begin{equation}
\hat{P}^J_{MK} | \Phi \rangle = c_{JK} | J, K \rightarrow M \rangle.
\end{equation}
Then one diagonalizes the Hamiltonian in the truncated space of
\begin{equation}
\left\{ \hat{P}^J_{M,-J} | \Phi \rangle, ~~ \hat{P}^J_{M,-J+1} | \Phi \rangle,~~ \cdots,~~ \hat{P}^J_{M,J} | \Phi \rangle \right\}
\end{equation}
Because the Hamiltonian is a scalar operator, thus commuting with the projection operator, and because $(\hat{P}^J_{MK})^\dagger = \hat{P}^J_{KM}$, $\hat{P}^J_{K^\prime M} \hat{P}^J_{MK} = \hat{P}^J_{K^\prime K}$, we only need
\begin{eqnarray}
%\langle (\hat{P}^J_{ML} \Phi) | \hat{H} | \hat{P}^J_{MK} \Phi \rangle = 
\langle \Phi | \hat{H} \hat{P}^J_{MK} | \Phi \rangle \equiv {\mathcal H}^J_{MK}, \\
%\langle (\hat{P}^J_{ML} \Phi) | \hat{P}^J_{MK} \Phi \rangle  = 
\langle \Phi | \hat{P}^J_{MK} | \Phi \rangle \equiv {\mathcal N}^J_{MK}.
\end{eqnarray}
We then solve a generalized eigenvalues problem, i.e., the  Hill-Wheeler equation, 
\begin{equation}
%\forall K', 
\sum_K {\mathcal H}^J_{K'K} g^r_{JK} = \epsilon_{r, J} \sum_K {\mathcal N}^J_{K' K} g^r_{JK}. \label{eqn:HillWheeler}
\end{equation}
where $r$ denotes different states with angular momentum $J$, at most $2J+1$ states, and $g^r_{JK}$ values determine the projected wavefunction.

To project out states, either by quadrature or by LAP, one needs to rotate states about the Euler angles $\alpha, \beta, \gamma$, using the rotation operator
\begin{equation}
\hat{R}(\alpha, \beta, \gamma) = \exp \left (  \frac{i\gamma}{\hbar} \hat{J}_z \right )
\exp \left (  \frac{i\beta}{\hbar} \hat{J}_y \right ) \exp \left (  \frac{i\alpha}{\hbar} \hat{J}_z \right ).
\end{equation}
The matrix elements of the rotation operator between spherical tensors, that is, states with good total angular momentum $j$ and $z$-component $m$, is the Wigner-$D$ matrix~\cite{Edmonds1957},
$\langle j^\prime m^\prime | \hat{R}(\alpha, \beta, \gamma) | j m \rangle = \delta_{j^\prime j} D^j_{m^\prime m}(\alpha \beta \gamma).$

For a given set of Euler angles $\Omega = (\alpha, \beta, \gamma)$, 
 the matrix element of rotational operator $\hat{R}(\Omega)$ is
\begin{equation}
\langle \Phi | \hat{R}(\Omega) | \Phi \rangle = \sum_{I M K } c^*_{IK} c_{IM} D^I_{KM} (\Omega) \langle I K | I M \rightarrow K \rangle = \sum_{IMK} D^I_{KM}(\Omega) \mathcal{N}^I_{KM}.
\label{eqn:LAP}
\end{equation}
For a series of Euler angles $\Omega_1, \Omega_2, \cdots$, we calculate the  norm kernels $\langle \Phi | \hat{R}(\Omega_i) | \Phi \rangle$, as well as the $D$ functions $D^I_{KM}(\Omega_i)$,
% with $I \leq J_{max}, K, M = -I, \cdots, I$, with $J_{max}$ properly set,  
and then find $\mathcal{N}^I_{KM}$ as solutions to a set of linear equations.
Similarly the Hamiltonian kernel $\mathcal{H}^I_{KM}$ can be also found.

Traditional projection procedure relies on the orthogonality of the Wigner-$D$ functions and thus upon quadrature; therefore the more abscissas ( Euler angles) one takes, the more accurate the projected energies.
The basis for LAP,  Eq.~(\ref{eqn:LAP}) is exact, but in practice one truncates the sum over $I$ with $I \leq J_{max}$. To find $J_{max}$, one computes $f_J$, the fraction of $J$-components defined in (\ref{eqn:fJ}), with $f_J = \sum_M | c_{JM} |^2$, and sets $J_{max}$ such that $f_J \geq f_{tol}$ for all $J \leq J_{max}$. 
We follow \cite{2019Convergence} and take $f_{tol} \sim 10^{-4}$ typically.
The error in projection is due to this truncation.
As a result of finite $J_{max}$, in LAP the number of Euler angles is smaller.
We implemented both methods, and found that LAP is more than 10 times faster than quadrature, with same accuracy of the energies of low-lying states.

To carry out this projection we need to be able to rotate pair condensates. This is almost as straightforward as for Slater determinants in angular-momentum projected 
Hartree-Fock~\cite{PhysRevC.96.064304}: one simply rotates the single-particle states, 
defining the matrix element of the rotation operator on single-particle bases $\alpha, \beta$ as  %$\mathcal D_{\alpha \beta}$, 
\begin{equation}
\mathcal{D}_{\alpha \beta} (\Omega)
= \langle n_\alpha l_\alpha j_\alpha m_\alpha | \hat{R} (\Omega) | n_\beta l_\beta j_\beta m_\beta \rangle 
= \delta_{n_\alpha n_\beta } \delta_{ l_\alpha l_\beta } \delta_{ j_\alpha j_\beta } D^{j_\alpha}_{m_\alpha m_\beta }(\Omega).
\end{equation}
Under rotation, a nucleon-pair (\ref{eqn-pair-condensate}) then becomes
\begin{equation}
\hat{R}(\Omega) \hat{A}^\dagger = \frac{1}{2} \sum_{\alpha \beta \gamma \delta} A_{\alpha \beta} \mathcal{D}_{\gamma \alpha}(\Omega) \mathcal{D}_{\delta \beta}(\Omega) \hat{c}_\gamma^\dagger \hat{c}_\delta^\dagger
 = \frac{1}{2} \sum_{\gamma \delta} {A'}_{\gamma \delta} \hat{c}_\gamma^\dagger \hat{c}^\dagger_\delta
 \equiv \hat{A'}^\dagger,
\end{equation}
with new pair coefficients
\begin{equation}
A'_{\gamma \delta} = \sum_{\alpha \beta} A_{\alpha \beta} \mathcal{D}_{\gamma \alpha}(\Omega) \mathcal{D}_{ \delta \beta }(\Omega) = ( \mathcal{D}(\Omega) A \mathcal{D}(\Omega) ^\top)_{\gamma \delta}.
\end{equation}
Therefore, the matrix element of rotational operator $\hat{R}(\Omega)$ on a pair condensate is
\begin{equation}
\langle \hat{A}^n | \hat{R}(\Omega) | (\hat{A}^\dagger)^n \rangle = 
\langle \hat{A}^n | (\hat{A'}^\dagger)^n | 0 \rangle, ~~~ A' = \mathcal{D}(\Omega) A \mathcal{D}(\Omega)^\top,
\end{equation}
and similarly
\begin{equation}
\langle \hat{A}^n | \hat{H} \hat{R}(\Omega) | (\hat{A}^\dagger)^n \rangle = 
\langle \hat{A}^n | \hat{H} (\hat{A'}^\dagger)^n | 0 \rangle, ~~~ A' = \mathcal{D}(\Omega) A \mathcal{D}(\Omega)^\top.
\end{equation}

Parity projection is similar and straightforward, with a given wavefunction $|\Psi \rangle$, the positive and negative parity components are ($\hat{\mathcal P}$ is the space inversion operator)
\begin{equation}
\frac{1}{2}(1 \pm \hat{\mathcal P}) | \Psi \rangle.
\end{equation}
To invert a many-body wavefunction, is equivalent to inverting all its single particle orbits.
When a pair is operated on by the space inversion operator, an additional phase factor is generated for the pair structure coefficients,
\begin{equation}
\hat{\mathcal{P}} \hat{A}^\dagger = \frac{1}{2}\sum_{\alpha \beta} A_{\alpha \beta} (\hat{\mathcal{P}} \hat{c}^\dagger_\alpha) (\hat{\mathcal{P}} \hat{c}^\dagger_\beta)
= \frac{1}{2}\sum_{\alpha \beta} A_{\alpha \beta} (-1)^{l_\alpha + l_\beta} \hat{c}^\dagger_\alpha \hat{c}^\dagger_\beta.
\end{equation}
In this work, we do both angular momentum and parity projections for the results in $2s1d0g_{7/2}0h_{11/2}$ shell.

\subsection{Tensor decomposition of a pair condensate} \label{subsec-decomposition}

\begin{figure}
\centering
\includegraphics[width=0.7\linewidth]{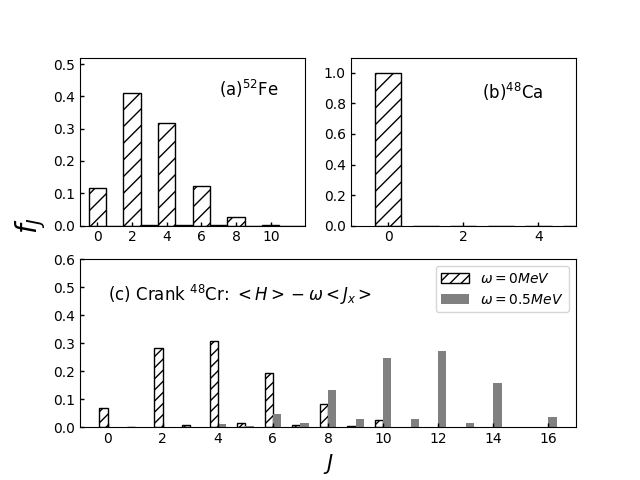}
\caption{(a) Fractions of $J$-components in the optimized pair condensate for $^{52}$Fe, as an example of rotational nuclei; (b) Fractions for $^{48}$Ca as an example of spherical nuclei; (c) Fractions for $^{48}$Cr as an example of cranking. All these cases are computed with the GX1A interactions in the $1p0f$ shell. }
\label{fig:fJ}
\end{figure}

As expressed by Eq.~(\ref{eqn-tensor-decomposition}), a pair condensate can be decomposed into components with different angular momentum quantum numbers $(J,M)$.
In other words, an optimized pair condensate is considered as a mixture of low-lying nuclear states.
We define the fraction of probability $f_J$ for total angular momentum $J$ in a pair condensate, as in \cite{PhysRevC.66.044318},
\begin{equation}
f_J = \sum_M |c_{JM} |^2, \label{eqn:fJ}
\end{equation}
where $c_{JM}$ is the amplitude of $(J,M)$ tensor component in the pair condensate, as in Eq.~(\ref{eqn-tensor-decomposition}).
To illustrate the decompositions for rotational  and spherical nuclei,  we show $^{52}$Fe and $^{48}$Ca, respectively, in Fig. \ref{fig:fJ}.
For $^{52}$Fe, in Fig. \ref{fig:fJ}(a) the decomposition is distributed across $J=0,2,4,6$ as expected for rotational nuclei. Conversely,
 for $^{48}$Ca in Fig. \ref{fig:fJ}(b), the wave function is almost entirely contained in the $J=0$ component; while 
there are  non-zero $f_J$ for $J >0 $ values, the magnitudes are too small to be seen in Fig. \ref{fig:fJ}(b).
Therefore the optimized condensate for semi-magic nuclei is mostly the spherical $J=0$ ground state, while tiny non-spherical components of the condensate are projected out to be the non-zero spectrum, which turn out to be surprisingly accurate in comparison with shell model results, as will be shown in Fig. \ref{fig:N26N28.isotones}.

However, if cranking, or variation with constraints, is involved, one can increase the fraction of components with higher $J$.
As in the usual procedure~ \cite{Ring1980}, one optimizes the trial wave function using the effective Hamiltonian 
\begin{equation}
\hat{H'} = \hat{H} - \omega \langle J_x \rangle.
\end{equation}
 Fig. \ref{fig:fJ}(c) shows  the decomposition of $^{48}$Ca at $\omega=0$ and $\omega=0.5$ MeV; the latter increases the components with $J>0$. 
Nonetheless for simplicity  we restrict ourselves to optimization without cranking for the rest of this paper.

\subsection{Electromagnetic Transitions}\label{subsec-EMtrans}

In this section we consider electromagnetic transitions between projected wave functions. As this is addressed in detail elsewhere, e.g. \cite{R.Rodriguez-Guzman2002}, 
 we briefly review the formalism.
Consider a state $|\Phi\rangle$, such as the optimized pair condensate; after diagonalization in the subspace of kets for angular momentum $J$,
\begin{equation}
\hat{P}^J_{MK} | \Phi \rangle, K = -J, -J+1, \cdots, J,
\end{equation}
the eigenstate is
\begin{equation}
\psi^r_{JM} = \sum_K g^r_{JK} \hat{P}^J_{MK} | \Phi \rangle = \sum_K g^r_{JK} | \hat{P}^J_{MK}\Phi \rangle 
\end{equation}
where $g^r_{JK}$ is the solution to the Hill-Wheeler equation (\ref{eqn:HillWheeler}),  and $| \hat{P}^J_{MK} \Phi \rangle =  \hat{P}^J_{MK} |\Phi \rangle $.
The one-body electromagnetic operator is denoted as (with proton and neutron parts),
\begin{equation}
\hat{Q}^s_\sigma = \hat{Q}^{\pi s}_\sigma + \hat{Q}^{\nu s}_\sigma,
\end{equation}
where $s$ is the angular momentum rank of the operator and $\sigma$ the $z$-component. 
The matrix element of the transition operator on projected kets is,
\begin{equation}
\langle \Phi | (\hat{P}^{J'}_{M'K'})^\dagger Q^s_\sigma \hat{P}^J_{MK} | \Phi \rangle = (JM,s\sigma|J'M') \langle (\hat{P}^{J'}_{*K'} \Phi) || Q^s || \hat{P}^J_{* K} \Phi \rangle,
\end{equation}
where $\langle (\hat{P}^{J'}_{*K'} \Phi) || Q^s || \hat{P}^J_{* K} \Phi \rangle$ is the reduced matrix element and $(JM,s\sigma|J'M')$ the Clebsch-Gordan coefficient~\cite{Edmonds1957}. 
As $(\hat{P}^{J'}_{M' K'})^\dagger = \hat{P}^{J'}_{K' M'}$, and the projection operator $\hat{P}^{J'}_{K' M'}$ just picks out the $(J',M')$ component and rotates it to $(J',K')$,
\begin{eqnarray}
\langle \Phi | (\hat{P}^{J'}_{M'K'})^\dagger Q^s_\sigma \hat{P}^J_{MK} | \Phi \rangle
&=& \langle \Phi | (\hat{P}^{J'}_{M'K'})^\dagger \sum_{J'' M''} (s \sigma, J M | J'' M'')(Q^s \hat{P}^J_{*K}|\Phi)^{J''}_{M''} \rangle, \nonumber \\
&=& (s \sigma, J M | J' M' ) \langle \Phi | (Q^s \hat{P}^J_{*K} )^{J'}_{K'} | \Phi \rangle.
\end{eqnarray}
Therefore the reduced matrix element can be computed by
\begin{eqnarray}
\langle (\hat{P}^{J'}_{*K'} \Phi) || Q^s || \hat{P}^J_{* K} \Phi \rangle &=& (-1)^{s + J - J'} \langle \Phi | (Q^s \hat{P}^J_{*K} )^{J'}_{K'} | \Phi \rangle \nonumber \\
&=& \sum_{\sigma} (J, K'-\sigma; s, \sigma | J' K') \langle \Phi | Q^s_\sigma \hat{P}^J_{K'-\sigma, K} | \Phi \rangle. \label{RME_Q}
\end{eqnarray}
%With given $J', K', J, K, s$, one needs to traverse $\sigma = -s, \cdots, s$, and calculate all the quantities $\langle \Phi | Q^s_\sigma \hat{P}^J_{MK} | \Phi \rangle$, and use these quantities to calculate (\ref{RME_Q}).
%In this way, there is only one projection operator other than two in $\langle \Phi | Q^s_\sigma \hat{P}^J_{MK} | \Phi \rangle$ and the task is lighter.
Here the needed projections are also carried out using the linear algebra  method.

Between the initial and final eigenfunctions $\psi^r_{JM} = \sum_K g^r_{JK} \hat{P}^J_{MK} | \Phi \rangle$, % and the final eigenfunciont $\psi^{r'}_{J'M'} = \sum_{K'} g^{r'}_{J'K'} \hat{P}^{J'}_{M'K'} |\Phi \rangle$, 
the reduced matrix element of $\hat{Q}^s$ is then finally
\begin{eqnarray}
\langle \psi^{r'}_{J'} || \hat{Q}^s || \psi^r_J \rangle &=&
\sum_{KK'} g^r_{JK} g^{r'}_{J'K'} \langle (\hat{P}^{J'}_{* K'} \Phi ) || \hat{Q}^s || ( \hat{P}^J_{* K} \Phi ) \rangle \nonumber\\
&=& \sum_{KK'} g^r_{JK} g^{r'}_{J'K'} \sum_{\sigma M} (JM,s\sigma|J'K') \langle \Phi | Q^s_\sigma \hat{P}^J_{MK} | \Phi \rangle.
\end{eqnarray}
The reduced probability, or  B-value, is
\begin{eqnarray}
B(F,J_r \rightarrow J'_{r'} ) &=& \frac{2J' +1}{2J +1} | \langle \psi^{r'}_{J'} || \hat{Q}^s || \psi^r_J \rangle |^2 \nonumber\\
&=& \frac{2J' +1}{2J+1} \left| \sum_{K K' \sigma }  g^r_{JK} g^{r'}_{J'K'} (J \, K'-\sigma, s \,\sigma | J' K') \langle \Phi | Q^s_\sigma \hat{P}^J_{K'-\sigma, K} | \Phi \rangle \right|^2.
\end{eqnarray}

\section{Benchmark results and Analysis}\label{sec-results}
In section \ref{subsec:sd}, we present ground state energies of all $1s0d$-shell even-even nuclei, with comparisons against full configuration-interaction shell model and projected Hartree-Fock calculations, and show ground band spectra as well as electric quadrupole transitions strengths, or B(E2) values, of silicon isotopes as an example.
In section \ref{subsec:pf}, we show $1p0f$-shell results, including shape evolution shown in lowest spectrum of $^{46,48}$Ca, $^{48,50}$Ti, $^{50,52}$Cr and backbending phenomenon in $^{52}$Fe.
Finally in section \ref{subsec:sdg7h11}, we illustrate the applicability to medium-heavy nuclides  with  $^{104}$Sn, $^{106}$Te, $^{108,124,126}$Xe, and $^{126,128}$Ba as examples in the $2s1d0g_{7/2}0h_{11/2}$ shell. 

\subsection{The $1s0d$ shell} \label{subsec:sd}

\begin{figure}
\centering
\includegraphics[width=1\linewidth]{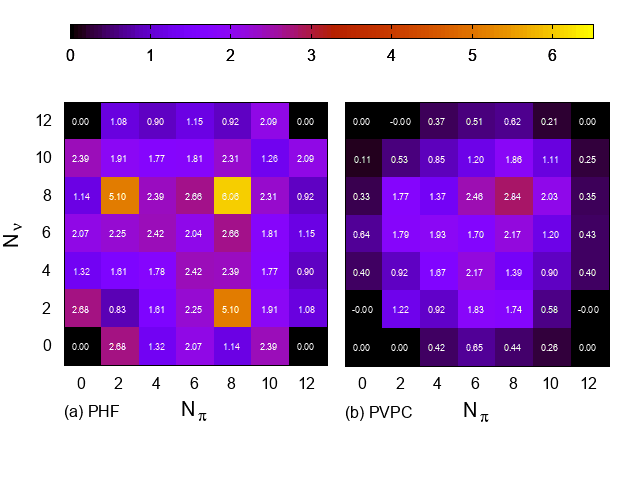}
\caption{ Ground state energy differences in MeV between approximate calculations and ``exact diagonalization" from full configuration-interaction shell model, $\Delta E^\mathrm{approx}_\mathrm{g.s} = E^\mathrm{approx}_\mathrm{g.s.} - E^\mathrm{SM}_\mathrm{g.s.}$, for all even-even nuclei in $1s0d$ shell. Panel (a) is for angular momentum-projected Hartree Fock results, and Panel (b) is for the PVPC in this work. See text for more details. }
\label{fig:sd.DeltaEgs}
\end{figure}

In variational methods, the ground state energy is often used as a marker of the quality of the approximation.
Because the $1s0d$ shell computations are easy and fast for shell model codes, we compute ground state energies of all even-even nuclei in the $1s0d$ shell, 
comparing PVPC against full configuration interaction shell model diagonalization, using the {\sc bigstick} code \cite{2013Factorization} and angular-momentum projected Hartree-Fock (PHF) calculations, e.g. \cite{lauber2021benchmarking},  using the exact same shell-model space and interaction input.
In particular we characterize the discrepancy in the ground state energy relative to the full configuration shell model (SM)  results, which for us are ``exact'', by
\begin{equation}
\Delta E^\mathrm{approx}_\mathrm{g.s.} = E^\mathrm{approx}_\mathrm{g.s.} - E^\mathrm{SM}_\mathrm{g.s.}.
\end{equation}
Here ``approx'' can be either PHF or PVPC. 
In Fig. \ref{fig:sd.DeltaEgs}, $\Delta E^\mathrm{PHF}_\mathrm{g.s.}$ and $\Delta E^\mathrm{PVPC}_\mathrm{g.s.}$ are shown as so-called `heat maps' indexed by colors. 
%Because we used shell-model interactions with isospin as an exact symmetry, isospin mirrors have the exact same ground state energy (modulo any Coulomb shift) and excitation spectra.
For most cases, the PHF ground state energy is about 2 MeV higher than the ``exact" ground state energy.
For three nuclei, $\Delta E^\mathrm{PHF}_\mathrm{g.s.}$ is more than 5 MeV.
For example, $^{32}$S is a rotational nucleus, while the Hartree-Fock minimum generates a spherical shape for it, therefore only a $0^+$ is projected out, as is also pointed out in \cite{Johnson_2020}.
PVPC is rather accurate for ground states of semi-magic nuclei, with $\Delta E^\mathrm{PVPC}_\mathrm{g.s.}$ typically less than $0.5$ MeV, which is unsurprising, as generalized seniority-0 configurations are contained within the pair condensate ansatz.
For rotational nuclei, $\Delta E^\mathrm{PVPC}_\mathrm{g.s.}$ is typically $1\sim 2$ MeV, i.e. PVPC has ground states of rotational nuclei slightly lower than that of PHF.
These are benchmark results for one reference state, i.e. one pair condensate for PVPC or one Slater determinant for PHF.
We expect inclusion of more reference states as done in generator-coordinate calculations~\cite{J.M.Yao2010} will reduce  discrepancies.

\begin{figure}
\centering
\includegraphics[width=0.7\linewidth]{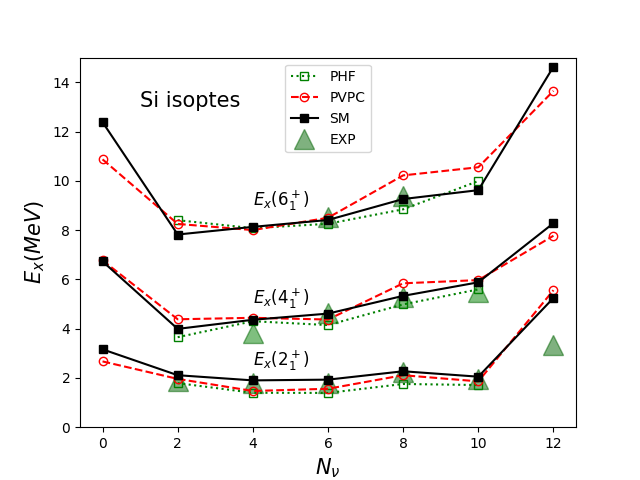}
\caption{Excitation energy of yrast states of even-even nuclei $^{22-34}$Si. "PHF" stands for angular momentum projected Hartree-Fock, "PVPC" stands for angular momentum projection after variation of a pair condensate (this work), "SM" stands for the full configuration-interaction shell model. 
All these three methods use the same occupation space $1s0d$, and the same interaction: USDB\cite{PhysRevC.74.034315}.  Experimental data are  from \cite{NSR2002KA80,BASUNIA20161,SHAMSUZZOHABASUNIA20131189,SHAMSUZZOHABASUNIA20102331,NSR1986DU07,NICA20121563}. }
\label{fig:Si.isotopes.spectrum}
\end{figure}

Aside from the ground state energies, it is interesting to see how well the excitation energies are approximated by such ``projection after variation" methods, as 
was done recently for PHF~\cite{lauber2021benchmarking}. 
As a test we investigate yrast bands of even-even isotopes of silicon, using results from PVPC, PHF and full configuration-interaction shell model.
Fig. \ref{fig:Si.isotopes.spectrum} demonstrates both PHF and PVPC generate good yrast-band excitation spectra for open-shell nuclei.
For semi-magic nuclei $^{22,34}$Si,  Hartree-Fock minima are  spherical and thus  PHF trivially generates only the ground state $0^+_1$.
The PVPC generates yrast bands of $^{22,34}$Si close to that of the shell model. 
For $^{34}$Si both the PVPC and the shell model predict the $2^+_1$ excitation energy higher than the experimental value~\cite{NICA20121563}.
We note that approximate excitation energies do not follow a variational principle and can be lower than that of the full configuration shell model.
% because approximate $0^+_1$ ground states energies are higher than  that of the shell model.
%When projecting from one reference, the number of $J^+$ states is $(2J+1)$ or less.
Although it is not shown here, the yrare and higher bands of PVPC can be worse, and likely need more reference states to improve, such as from coexisting local 
minima~\cite{lauber2021benchmarking}. 

As a test of the approximate wavefunctions, we present in Table \ref{tab:sdBE2} the $B(E2, I+2 \rightarrow I)$ values for silicon isotopes corresponding to Fig. \ref{fig:Si.isotopes.spectrum}.
While projected states can reproduce the transition strengths qualitatively, there can still be discrepancies, for example $B(E2, 4^+_1 \rightarrow 2^+_1)$ of $^{26}$Si from the PVPC is about $ 1/6$ of that from the shell model; again, this may be improved by additional reference condensates.

\begin{table}
\centering
\begin{tabular}{ccccc}
\hline \hline
Nuclide & Method & $2^+_1 \rightarrow 0^+_1$ &  $4^+_1 \rightarrow 2^+_1$ &  $6^+_1 \rightarrow 4^+_1$  \\ 
\hline
\multirow{2}{*}{$^{22}$Si} & SM & 33.32 & 16.96 & 13.07 \\ 
 & PVPC & 32.28  & 14.10  & 14.81  \\ 
\hline 
\multirow{2}{*}{$^{24}$Si} & SM & 44.19 & 23.89 & 17.58 \\ 
 & PVPC & 37.13  & 25.60  & 41.66  \\ 
\hline 
\multirow{2}{*}{$^{26}$Si} & SM & 44.64 & 11.29 & 41.94 \\ 
 & PVPC & 37.34  & 60.31  & 70.56  \\ 
\hline 
\multirow{2}{*}{$^{28}$Si} & SM & 78.48 & 110.4 & 94.71 \\ 
 & PVPC & 83.47  & 116.3  & 119.2  \\ 
\hline
\multirow{2}{*}{$^{30}$Si} & SM & 46.52 & 15.94 & 32.70 \\ 
 & PVPC & 48.33  & 33.68  & 78.15  \\ 
\hline
\multirow{2}{*}{$^{32}$Si} & SM & 43.52 & 67.11 & 52.70 \\ 
 & PVPC & 37.54  & 64.06  & 66.85  \\ 
\hline
\multirow{2}{*}{$^{34}$Si} & SM & 34.91 & 16.80 & 6.14 \\ 
 & PVPC & 30.78  & 14.93  & 1.50  \\ 
\hline\hline
\end{tabular}
\caption{$B(E2)$ values for silicon isotopes in units of $e^2\, \mathrm{fm}^4$. SM stands for full configuration shell model with ``exact" diagonalization, while PVPC stands for projection and variation of one pair condensate. Both methods use the USDB interaction in the $1s0d$ shell. Effective charges are set as $e_p=1.5e, e_n=0.5e$. \label{tab:sdBE2}}
\end{table}

\subsection{The $1p0f$ shell} \label{subsec:pf}
We use the GX1A interaction in  the $1p0f$ shell \cite{PhysRevC.65.061301, Honma2005}.
We find the PVPC ground state energy is typically  $1$-$2$ MeV above the exact shell model value.
Here we give results of Ca, Ti, Cr isotones with $N=26,28$, to show that PVPC generates spectrum of pairing-like and vibrational-like spectrum self-consistently, as the number of valence particles increase.
Then we demonstrate backbending  in the rotational yrast spectrum of  $^{52}$Fe, with the pair condensate as the intrinsic wave function.

\subsubsection{$^{46,48}${\rm Ca}, $^{48,50}${\rm Ti}, $^{50,52}${\rm Cr} } 
\begin{figure}
\centering
\includegraphics[width=1.1\linewidth]{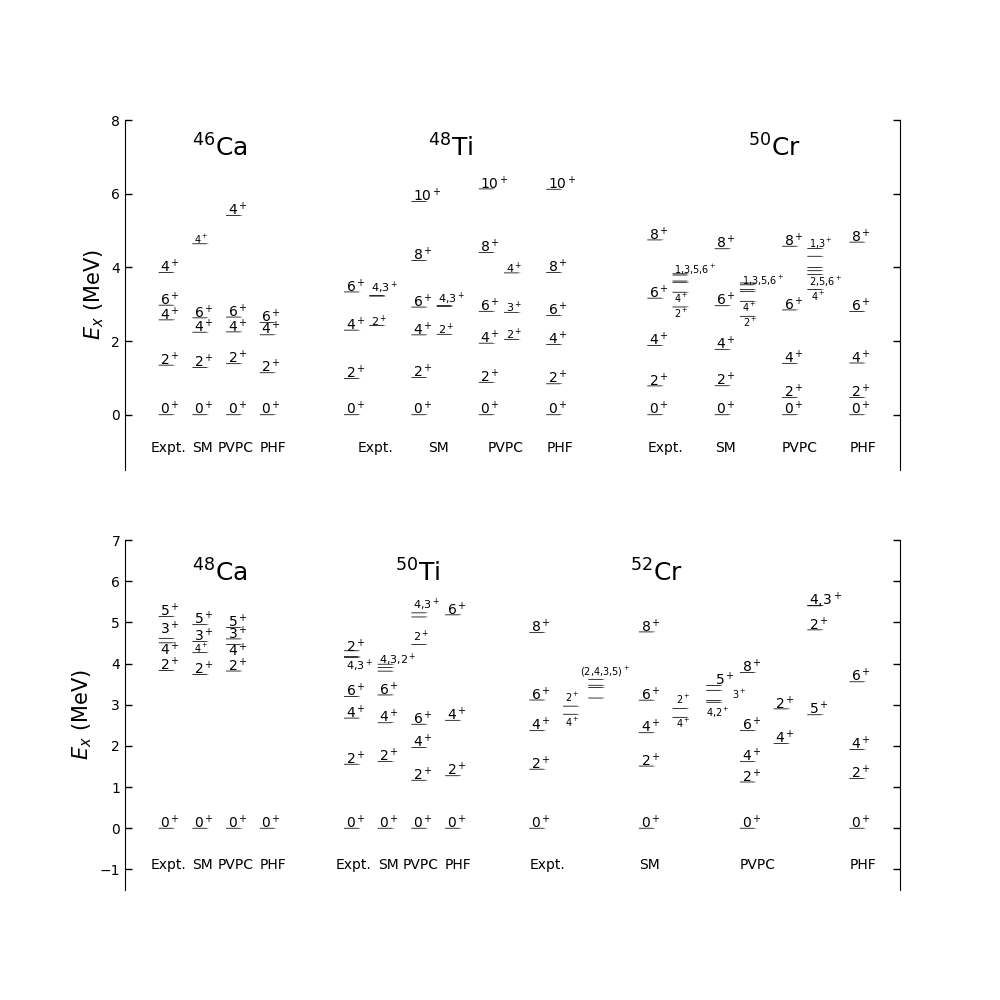}
\caption{ Low-lying spectrum of $N=26,28$ isotones: $^{46,48}$Ca, $^{48,50}$Ti, $^{50,52}$Cr. All theoretical results are produced with the  interaction GX1A \cite{PhysRevC.65.061301,Honma2005} in the $1p0f$ shell. Experimental results are  from \cite{WU20001,BURROWS20061747,CHEN20191,DONG2015185}.}
% The PVPC and PHF results each have one reference state. }
\label{fig:N26N28.isotones}
\end{figure}

 Fig. \ref{fig:N26N28.isotones} presents the low-lying spectra of $N=26,28$ isotones $^{46,48}$Ca, $^{48,50}$Ti, $^{50,52}$Cr, using results from PVPC, PHF,  and the 
full configuration shell model, as well as comparison with experiment which illustrates the quality of the shell model interaction. 
%We note that PVPC and PHF generate at most $2J+1$ states for $J^+$, therefore we only show corresponding states from shell model for comparison.
For the spherical nucleus $^{48}$Ca, the fraction $f_J$ for $J=0$ is more than $99.9\%$, as shown in Fig. \ref{fig:fJ}; nonetheless, the $J > 0$  states projected out with  $f_J < 0.1\%$  turn out to be unexpectedly accurate. We see seniority-like spectra with a substantial pairing gap between $0^+_1$ and $2^+_1$, and the $3+,4+,5+$ levels clustered close to $2^+_1$.
Although it is not shown on Fig. \ref{fig:N26N28.isotones}, the $0^+_1$ of $^{46,48}$Ca is  $\approx 0.3$ MeV above the ``exact" shell model ground state, while other nuclei in the picture have ground state energies $1\sim 2$ MeV above shell model results.
The excitation spectra of the ``ground band" from all three theoretical methods agree qualitatively, except that PHF does not generate $J > 0$ states for $^{48}$Ca, with a spherical Hartree-Fock energy about $0.4$ MeV above the exact shell model value.

Table~\ref{tab:pfBE2} provides the corresponding $B(E2)$ values for the nuclides in Fig. \ref{fig:N26N28.isotones}. The agreement between PVPC and the full configuration shell model supports the quality of PVPC wavefunctions, although we note that $B(E2;6^+_1 \rightarrow 4^+_1)$ of $^{48}$Ca by PVPC is far from the shell model result. 
We note that the shell model spectra  of $^{48}$Ca has four $6^+$ states with an excitation energy between 7.7 and 9 MeV,
while  PVPC yields only one $6^+$ state in the same energy range. Therefore the $6^+_1$ by PVPC is likely  a mixture of those shell model $6^+$ states, a possible origin for the underestimation of $B(E2;6^+_1 \rightarrow 4^+_1)$.
\begin{table}
\centering
\begin{tabular}{ccccc|ccccccc}
\hline \hline
Nucleus & Method & $2^+_1 \rightarrow 0^+_1$ &  $4^+_1 \rightarrow 2^+_1$ &  $6^+_1 \rightarrow 4^+_1$ & Nucleus & Method & $2^+_1 \rightarrow 0^+_1$ &  $4^+_1 \rightarrow 2^+_1$ &  $6^+_1 \rightarrow 4^+_1$ \\%&  $8^+_1 \rightarrow 6^+_1$  \\ 
\hline
\multirow{2}{*}{$^{46}$Ca} & SM & 7.72 & 6.02 & 3.08 & \multirow{2}{*}{$^{48}$Ca} & SM & 10.12 & 1.99 & 4.03 \\
 & PVPC & 7.31  & 4.10  & 2.16  &  & PVPC & 9.10  & 1.76  & 0.36  \\  
\hline 
%& EXP & 35.54 &  & 5.38 & & EXP & 
\multirow{2}{*}{$^{48}$Ti} & SM & 87.89 & 126.06 & 51.28 & \multirow{2}{*}{$^{50}$Ti} & SM & 85.20 & 83.10 & 39.84 \\
 & PVPC & 84.67  & 116.36  & 113.57 &  & PVPC &  65.80  & 58.77  & 19.87  \\ 
\hline 
\multirow{2}{*}{$^{50}$Cr} & SM & 184.85 & 261.43  & 221.87 & \multirow{2}{*}{$^{52}$Cr} & SM & 142.31 & 91.94  & 45.91  \\ 
 & PVPC & 190.33 & 257.78  & 266.68  &  & PVPC & 102.8 & 89.00  & 56.39  \\ 
\hline \hline
%\multirow{2}{*}{$^{52}Fe$} & SM & 188.67 & 246.20 & 171.09 \\ 
% & PVPC &  158.68  & 212.19  & 185.70  \\ 
%\hline\hline
\end{tabular} 
\caption{$B(E2)$ value of $^{48}$Ca, $^{50}$Ti, $^{52}$Cr, $^{52}$Fe in $e^2fm^4$. The PVPC and the shell model both use the GX1A interactions \cite{PhysRevC.65.061301,Honma2005}, with effective charges $e_p = 1.5e, e_n = 0.5e$, and oscillator length $A^{1/6}$ fm. \label{tab:pfBE2}}
\end{table}

\subsubsection{Backbending  in $^{52}${\rm Fe} }

\begin{figure}
\centering
\includegraphics[width=0.7\linewidth]{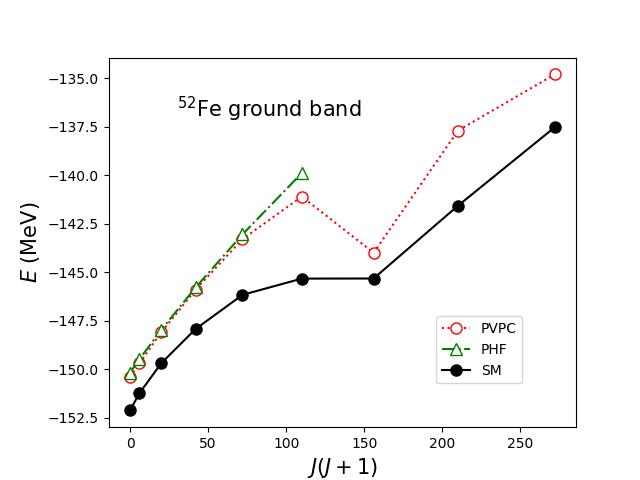}
\caption{Backbending of yrast band of $^{52}$Fe. ``PVPC'' is projected states from one pair condensate,  ``SM'' is the full configuration shell model, and ``PHF''  the projected Hartree-Fock method. The GX1A interaction is used for the $1p0f$ shell.
\label{fig:Fe52.backbending} }
\end{figure}

While many methods using angular momentum projection produce good rotational bands, e.g.  projected Hartree-Fock \cite{PhysRevC.96.064304} and  the projected shell model \cite{Sun2016}, a more stringent test is backbending.
%It is interesting to check the back-bending phenomena, and see how well a single pair condensate wears it out.
For a simple rotor picture, the excitation energy of a yrast state $E_x(J^+_1)$ is proportional to $J(J+1)$,  but both experiment 
and theory see abrupt changes in the moment of inertia and thus the yrast spectrum deviates from the  simple $J(J+1)$ rule.
One explanation of this is pair breaking due to the increasing Coriolis force~\cite{Ring1980}.
In Fig. \ref{fig:Fe52.backbending} we show yrast energies of $^{52}$Fe, from PVPC with a pair condensate playing the role of the intrinsic state, and that from exact diagonalization with the shell model.
The PVPC energies are about $1$ MeV or more above the full configuration shell model energies, but both PVPC and the shell model exhibits two backbends.
%But both methods exhibit two kinks.
The PVPC curve shows more abrupt changes than the shell model.
%In the $J=[0,12]$ zone, the shell model exhibits a smooth back-bending curve, while the PVPC curve is more stiff, though with a slight back-bending tendency, and a heavy dip down at $J=12, J(J+1)=156$.
%After $J=12$, the energies soar up again, which can be explained as a shell effect in the model space.
The change for $J > 12$ is likely because 6 valence protons and 6 valence neutrons restricted to the $0f_{7/2}$ orbit have a maximum angular momentum of 12.

%in the $1p0f$ shell, with a large probability occupying the $0f_{7/2}$ orbit at low energy, because $0f_{7/2}$ is lower, with a small subshell gap, below the other 3 orbits $1p_{3/2}, 0f_{5/2}, 1p_{1/2}$.
%As 6 protons or 6 neutrons in $0f_{7/2}$ has at most $J=6$ respectively, therefore total angular momentum coupled by them is at most $J=12$.
%Therefore, states with $J > 12$ have to count on configurations involving higher orbits $1p3/2, 0f5/2, 1p1/2$, which results in the energy gain at $J=14,16$ in Fig.\ref{fig:Fe52.backbending}.
Although we do not show it, cranked PHF also generates two kinks;
%We note that cranked PHF also generates such two kinks, although it is not shown in Fig. \ref{fig:Fe52.backbending}.
without cranking, the Slater determinant of the Hartree-Fock minimum does not have components of $J \geq 12$.
%By transforming to the canonical basis, as discussed in Subsec. \ref{subsec:sd}, the optimized pair condensate approximates one Slater determinant, but with tiny deviations which accounting for the $J \geq 12$ states of PVPC.

\subsection{The $2s1d0g_{7/2}0h_{11/2}$ shell}\label{subsec:sdg7h11}

\begin{figure}
\centering
\includegraphics[width=0.9\linewidth]{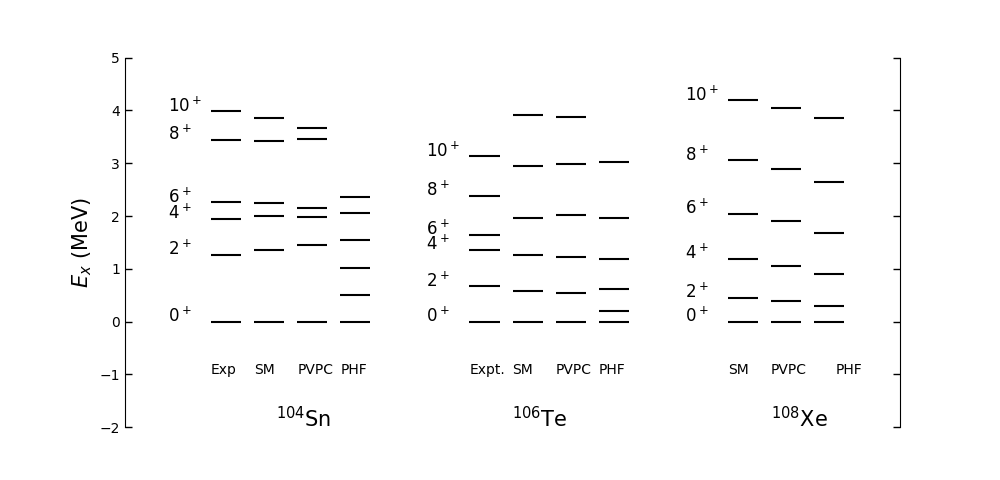}
\caption{ Ground band excitation energies of $N=54$ isotones $^{104}$Sn, $^{106}$Te, $^{108}$Xe. Both the shell model and the PVPC use a monopole-adjusted G-matrix interaction \cite{PhysRevC.86.044323}, and the experimental data is from \cite{BLACHOT20072035,DEFRENNE2008943}. }
\label{fig:N54}
\end{figure}

\begin{figure}
\centering
\includegraphics[width=1.1\linewidth]{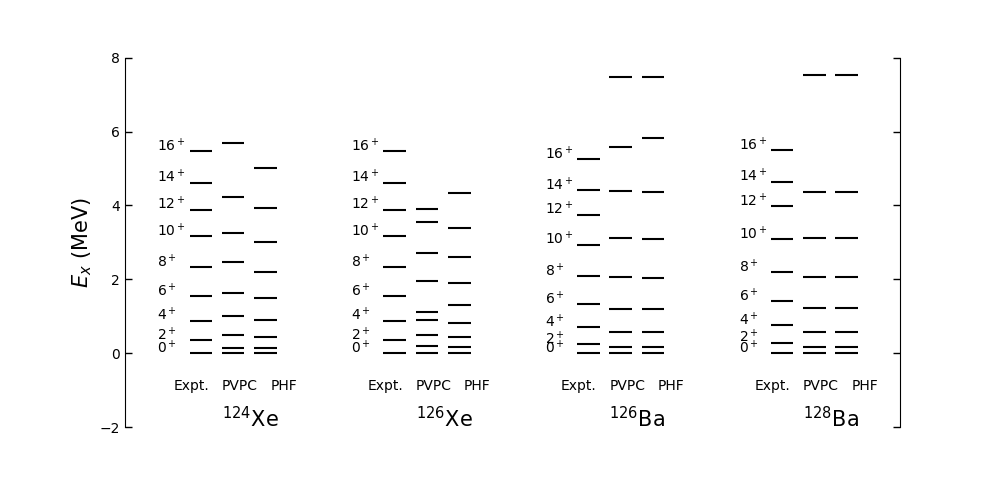}
\caption{ Ground band excitation energy of $^{124, 126}$Xe, $^{126,128}$Ba,  }
\label{fig:xeba}
\end{figure}

Fig. \ref{fig:N54} shows the ground state band of three $N=54$  isotones. 
For $^{104}$Sn, $^{106}$Te we take the experimental data  from \cite{BLACHOT20072035,DEFRENNE2008943}, while
the low-lying spectrum of $^{108}$Xe has not been not measured.
%The shell model dimension becomes too large soon as valence particles increase, therefore these isotopes with a few valence particles serve as benchmark examples for comparison among the shell model, PVPC and PHF.
PVPC generates ground band rather close to that of the shell model.
The discrepancy $\Delta E^{PVPC}_{g.s.} \equiv E^{PVPC}_{g.s.} - E^{SM}_{g.s.}$ is about $0.2, 0.8, 0.9$ MeV, respectively for $^{104}$Sn, $^{106}$Te, $^{108}$Xe.
As in the $1s0d$ or $1p0f$ shell, PVPC does better for semi-magic nuclei than PHF, as one might anticipate, but PHF approaches PVPC for the 
rotational nucleus $^{108}$Xe, although the PVPC ground state energy is about $0.5$ MeV lower than that of PHF.
In Table \ref{tab:N54BE2}, the $B(E2, I^+_1 \rightarrow (I-2)^+_1)$ values are presented, in correspondence to Fig. \ref{fig:N54}.
The $B(E2)$ values of PVPC agree with those of the shell model qualitatively.
We also show the ground band of $^{124, 126}$Xe, $^{126,128}$Ba in Fig. \ref{fig:xeba} generated with PVPC and PHF, although $0^+_1$ of both these methods appear not low enough.
%In computation, PVPC does not invoke the Pandya transformation, because the pair condensate form does not have particle-hole symmetry, i.e. a particle-pair condensate does not transform into a hole-pair condensate ( in our practice PVPC works better in the particle representation ).
%For example in $^{126}$Xe, we use a proton pair condensate with $N_\pi = 4$, and a neutron pair condensate with $N_\nu = 22$, to do variation and angular momentum projection.
For all these four nuclei $^{124,126}$Xe, $^{126,128}$Ba the PVPC generates ground state energies less than $0.5$ MeV lower than PHF does, and the excitation energies of PVPC and PHF are mostly alike.

\begin{table}
\centering
\begin{tabular}{ccccccc}
\hline \hline
Nucleus & Method & $2^+_1 \rightarrow 0^+_1$ &  $4^+_1 \rightarrow 2^+_1$ &  $6^+_1 \rightarrow 4^+_1$ & $8^+_1 \rightarrow 6^+_1$ & $10^+_1 \rightarrow 8^+_1$ \\ 
\hline
\multirow{2}{*}{$^{104}$Sn} & SM & 42.9 & 40.0 & 14.2 & 18.1 & 30.8 \\ 
 & PVPC & 36.1  &  24.5 & 5.4 &  10.8 & 19.6 \\ 
\hline
\multirow{2}{*}{$^{106}$Te} & SM & 444.8 & 588.0 & 573.0 & 478.6 & 504.8 \\ 
 & PVPC & 430.3  &  597.8 & 619.7 &  565.8 & 485.0 \\ 
\hline
\multirow{2}{*}{$^{108}$Xe} & SM & 943.3 & 1277 & 1373 & 1449 & 1369 \\ 
 & PVPC & 755.6  & 1044  & 963.2 & 901.1  & 1044 \\
\hline\hline
\end{tabular}
\caption{$B(E2)$ value of $N=54$ isotones $^{104}$Sn, $^{106}$Te, $^{108}$Xe in $e^2$ fm$^4$. 
%SM  while PVPC stands for projection and variation of one pair condensate in this work. 
Both methods use a monopole-modified G-matrix interaction \cite{PhysRevC.86.044323} in the $2s1d0g_{7/2}0h_{11/2}$ shell. Effective charges are set as $e_p=1.5e, e_n=0.5e$. \label{tab:N54BE2}}
\end{table}

\section{summary}\label{sec-sum}

In summary, we derived formulas for projection after variation of pair condensates(PVPC), implemented those formulas into codes, and present one-reference (i.e. one pair condensate for one nucleus) benchmark results.
%The variation of a pair condensate is time-consuming, as it takes several hours to optimize a pair condensate for $^{132}$Dy in the $2s1d0g_{7/2}0h_{11/2}$ space on a 44-core machine, though the tri-axial projection is much less heavy with the LAP technique.
The variation of a pair condensate is time-consuming, as it takes several hours to optimize a pair condensate for $^{132}$Dy in the $2s1d0g_{7/2}0h_{11/2}$ space on a 44-core machine. After finding the variational minimum, triaxial projection with the LAP technique is very quick.
Because the pair condensate as a variation ansatz is more general than the Hartree-Fock, or the Hartree-Fock-Bogolyubov method, the PVPC generates semi-magic or open-shell nuclei more consistently.
We benchmark results of nuclei in $1s0d, 1p0f, 2s1d0g_{7/2}0h_{11/2}$ shells, by comparing results of PVPC with PHF and shell model when possible.
For semi-magic nuclei the PVPC generates seniority-type spectrum, satisfactorily close to shell model results.
For rotational nuclei the PVPC generates rotational ground band, similar to results from the Projected Hartree-Fock method, though less than $1$ MeV lower, and with a few more states.

In future work, a benchmark comparison with the Hartree-Fock-Bogoliubov method can be interesting, as the HFB and PVPC both have intrinsic pairing structure.
PVPC allows for tri-axial deformation, but due to constraint of one reference (one pair condensate), tri-axial phenemonon, such as a low $0^+_2$ in $^{76}$Ge, is not reproduced yet.
In addition, inclusion of more than one pair condensate \`a la generator coordinate methods can be implemented.
It is also straight forward to extend the formalism to proton-neutron pair condensates, or odd valence protons/neutrons, or separate projection of proton/neutron parts as related to the scissors mode.

\acknowledgements

%We would like to thank useful discussions with Fang-Qi Chen(陈芳祁), Jie Meng(孟杰), Guanjian Fu(孟杰), and part of shell model results from Yixin Yu(俞一信).
We would like to thank useful discussions with Fang-Qi Chen, Jie Meng, Guan Jian Fu, and part of shell model results from Yi Xin Yu.
Yi Lu acknowledges support by the National Natural Science Foundation of China (Grant No. 11705100, 12175115), the Youth Innovations and Talents Project of Shandong Provincial Colleges, and Universities (Grant No. 201909118), Higher Educational Youth Innovation Science and Technology Program Shandong Province (Grant No. 2020KJJ004).
Yang Lei is grateful for the financial support of the Sichuan Science and Technology Program (Grant No. 2019JDRC0017), and the Doctoral Program of Southwest University of Science and Technology (Grant No. 18zx7147). 
This material is also based upon work (Calvin W. Johnson) supported by the U.S. Department of Energy, Office of Science, Office of Nuclear Physics, under Award Number  DE-FG02-03ER41272.
%\end{CJK*}

\appendix 
\section{Formulas for Hamiltonian matrix elements and their derivatives} \label{app:MatrixElements}

To carry out the variation and projection of a pair condensate state $|\Phi\rangle \equiv (\hat{A}^\dagger)^n |0\rangle$ we need a variety of matrix elements.
In particular we need the normalization $\langle \Phi | \Phi \rangle$, then expectation of the Hamiltonian of that state $\langle \Phi | \hat{H} | \Phi \rangle$, and the partial derivatives of these two quantities: $\partial \langle \Phi | \Phi \rangle / \partial A_{\alpha \beta}$ and $\partial \langle \Phi | \hat{H} | \Phi \rangle / \partial A_{\alpha \beta}$.
To do projection, we also need to rotate the condensate $\hat{R}(\Omega) (\hat{A}^\dagger)^n |0\rangle = (\hat{A'}^\dagger)^n |0\rangle$, with coefficient matrix $A' = {\mathcal D}(\Omega) A {\mathcal D}(\Omega)^\top$, and to calculate the norm and Hamiltonian kernels: $\langle \hat{A}^n | \hat{R}(\Omega) | (\hat{A}^\dagger)^n \rangle = \langle \hat{A}^n | (\hat{A'}^\dagger)^n \rangle$ and $\langle \hat{A}^n | \hat{H} \hat{R}(\Omega) | (\hat{A}^\dagger)^n \rangle = \langle \hat{A}^n | \hat{H} | (\hat{A'}^\dagger)^n \rangle $.
%Overall, aside from overlaps, the Hamiltonian matrix elements between condensates $\langle \hat{A}_1^n | \hat{H} | (\hat{A}_2^\dagger)^n \rangle$, and their derivatives $\partial \langle \hat{A}^n_1 | \hat{H} | (\hat{A}^\dagger_2)^n \rangle / \partial (A_1)_{\alpha \beta}$ are important in our formalism.
In this appendix we give the expressions for Hamiltonian matrix elements and their derivatives in terms of condensate overlaps including what we term impurity pairs.  In Appendix \ref{app-overlaps}, we derive all necessary overlap formulas of condensates, with or without ``impurity" pairs, completing  the necessary formalism.

\subsection{Hamiltonian matrix elements}
%The expectation value of the Hamiltonian on the normalized trial wave is sometimes called the average energy, $\langle \hat{H} \rangle$.
The Hamiltonian matrix element, $\langle (\hat{A}_1)^n | \hat{H} | (\hat{A}^\dagger_2)^n \rangle$, consists of of one-body parts $\langle (\hat{A}_1)^n | \hat{Q} | (\hat{A}^\dagger_2)^n \rangle$, such as single particle energies, and two-body parts $\langle (\hat{A}_1)^n | \hat{A}^\dagger_3 \hat{A}_4 | (\hat{A}^\dagger_2)^n \rangle$.
The one-body operator
$\hat{Q} \equiv \sum_{\alpha \beta} Q_{\alpha \beta} \hat{c}^\dagger_\alpha \hat{c}_\beta$ has structure coefficients $Q_{\alpha \beta}$. %and can be the numerator $\hat{n}_a, \hat{n}_b$ in Eq. (\ref{eqn-hamiltonian}).
Through commutation relations, the matrix element of a one-body operator can be expressed as an overlap,
\begin{equation}
\langle (\hat{A}_1)^n | \hat{Q} | (\hat{A}_2^\dagger)^n \rangle 
= n \langle (\hat{A}_1)^{n-1} [\hat{A}_1, \hat{Q}] | (\hat{A}_2^\dagger)^n \rangle
= n \langle (\hat{A}_1)^{n-1}\hat{A}_3 | (\hat{A}_2^\dagger)^n \rangle,
\label{eqn:one-body}
\end{equation}
where $\hat{A}_3 \equiv{[} \hat{A}_1, \hat{Q} {]}$, with pair structure coefficients $A_3 = A_1 Q + Q^\top A_1$.  We term 
$\hat{A}_3$  an ``impurity pair", as it replaces one $\hat{A}_1$ in the bra vector $\langle (\hat{A}_1)^n |$ to arrive at Eq.~(\ref{eqn:one-body}).
% and ends up $\langle (\hat{A}_1)^{n-1} \hat{A}_3 |$.

Similarly, the matrix element of a general two-body interaction $\hat{A}^\dagger_3 \hat{A}_4$ can be written as
\begin{eqnarray}
 \langle (\hat{A}_1)^n | \hat{A}^\dagger_3 \hat{A}_4 | ( \hat{A}_2^\dagger )^n \rangle
&=&  \langle \hat{A}^n_1 \hat{A}_4 | \hat{A}^\dagger_3 (\hat{A}^\dagger_2)^n \rangle
 + \langle \hat{A}^n_1 [ \hat{A}^\dagger_3,  \hat{A}_4 ] (\hat{A}^\dagger_2)^n \rangle
  \label{eqn:two-body} \\
&=& \langle \hat{A}^n_1 \hat{A}_4 | \hat{A}^\dagger_3 (\hat{A}^\dagger_2)^n \rangle
  + \frac{1}{2} \mathrm{tr} \, ( A_3 A_4 ) \langle \hat{A}^n_1 | (\hat{A}^\dagger_2)^n \rangle
  - n \langle \hat{A}^{n-1}_1 \hat{A}_5 | (\hat{A}^\dagger_2)^n \rangle, \nonumber
\end{eqnarray}
where $ \mathrm{tr} \,( A_3 A_4 )$ is the trace of the matrix product $A_3 A_4$, and \(A_5 =  A_1 A_3 A_4 + A_4 A_3 A_1 \).
The first term on the right of Eq. (\ref{eqn:two-body}), $ \langle \hat{A}^n_1 \hat{A}_4 | \hat{A}^\dagger_3 (\hat{A}^\dagger_2)^n \rangle$, 
with two impurity pairs, is the most complicated overlap to evaluate in Eq. (\ref{eqn:two-body}).

\subsection{Derivatives of Hamiltonian matrix elements}
To calculate derivatives  $\partial \langle \hat{A}^n_1 | \hat{H} | (\hat{A}^\dagger_2)^n \rangle / \partial (A_1)_{\alpha \beta}$ and $\partial \langle \hat{A}^n_1 | (\hat{A}^\dagger_2)^n \rangle / \partial (A_1)_{\alpha \beta}$, we borrow from Ref. \cite{mcsm-pair}  a general formalism for derivatives.
Let $\hat{O}$ be an operator, e.g. 1 for overlap or $\hat{H}$ for Hamiltonian matrix elements. The matrix element of $\hat{O}$ on two pair condensates is
\begin{equation}
\langle \hat{A}^n_1 | \hat{O} | ( \hat{A}^\dagger_2 )^n \rangle.
\end{equation}
Taking $\hat{A}_1 (\lambda) = \hat{A}_1 + \lambda \hat{\Gamma}$,  where $\lambda$ is small (and, we assume, real), yields to first order 
%An infinite small increment $\delta_\Gamma \hat{\Gamma} $ onto $\hat{A}_1$ leads to 
\begin{equation}
\langle (\hat{A}_1 + \lambda \hat{\Gamma} )^n | \hat{O} | ( \hat{A}^\dagger_2 )^n \rangle
= \langle \hat{A}^n_1 | \hat{O} | ( \hat{A}^\dagger_2 )^n \rangle + n \lambda \langle \hat{A}^{n-1}_1 \hat{\Gamma} | \hat{O} | (\hat{A}^\dagger_2)^n \rangle,
\end{equation}
hence
%Therefore the partial derivative on $\delta_\Gamma$ can be derived, for real $\delta_\Gamma$ it is,
\begin{equation}
\left. \frac{\partial}{\partial \lambda} \langle \hat{A}^n_1(\lambda)  | \hat{O} | ( \hat{A}^\dagger_2 )^n \rangle \right |_{\lambda= 0} 
= n \langle \hat{A}^{n-1}_1 \hat{\Gamma} | \hat{O} | (\hat{A}^\dagger_2)^{n} \rangle.
\label{eqn-derivative}
\end{equation}
Now $\hat{\Gamma}$ plays the role of an impurity pair.
%Therefore derivatives induce more ``inpurity pairs", for example, 
The derivative of an overlap between pair condensates is simply
\begin{eqnarray}
\frac{\partial}{\partial \lambda} 
 \langle (\hat{A}_1(\lambda) )^n | ( \hat{A}_2^\dagger )^n \rangle
= n \langle (\hat{A}_1)^{n-1} \Gamma | ( \hat{A}_2^\dagger )^n \rangle,
\end{eqnarray}
while the derivative of a one-body matrix element (Eq. \ref{eqn:one-body}) is
\begin{eqnarray}
\frac{\partial}{\partial \lambda} 
 \langle (\hat{A}_1)^n | \hat{Q} | ( \hat{A}_2^\dagger )^n \rangle
= n \langle (\hat{A}_1)^{n-1} \Gamma | [\hat{Q}, \hat{A}_2^\dagger] ( \hat{A}_2^\dagger )^{n-1} \rangle = n \langle (\hat{A}_1)^{n-1} \Gamma | \hat{A}^\dagger_3 ( \hat{A}_2^\dagger )^{n-1} \rangle,
\end{eqnarray}
where $\hat{A}^\dagger_3 \equiv [\hat{Q}, \hat{A}_2^\dagger]$, with pair structure coefficients $A_3 = A_2 Q^\top + Q A_2 $.
Finally, the derivative of two-body matrix element (Eq. \ref{eqn:two-body}) is
\begin{eqnarray}
\frac{\partial}{\partial \lambda} 
 \langle (\hat{A}_1)^n | \hat{A}^\dagger_3 \hat{A}_4 | ( \hat{A}_2^\dagger )^n \rangle
&=&  n \langle \hat{A}^{n-1}_1 \hat{\Gamma} | \hat{A}^\dagger_3 \hat{A}_4 | (\hat{A}^\dagger_2)^{n} \rangle 
\nonumber \\
&=& n \langle \hat{A}^{n-1}_1 \hat{\Gamma} \hat{A}_4 | \hat{A}^\dagger_3 (\hat{A}^\dagger_2)^n \rangle
+  n \, \mathrm{tr}\, (A_3 A_4) \langle \hat{A}_1^{n-1} \hat{ \Gamma } | ( \hat{A}^\dagger_2 )^n \rangle \nonumber\\
&& - n(n-1) \langle \hat{A}^{n-2}_1 \Gamma \hat{A}_5 | (\hat{A}^\dagger_2)^n \rangle,
\label{eqn:deriv-two-body}
\end{eqnarray}
where $A_5 = A_1 A_3 A_4 + A_4 A_3 A_1 $.
On the right side, the most complicated overlap in this work is the first term, with 3 impurity pairs.
Thus the  derivatives of Hamiltonian matrix elements can be computed from overlaps of pair condensates with impurity pairs, discussed in Appendix \ref{app-overlaps}.
%All in all, the overlap formulas in Appendix \ref{app-overlaps} are the corner stones of the whole formalism for variation and angular momentum projection.

\section{Overlap formulas}\label{app-overlaps}

In this appendix, we give formulas for overlaps between pair condensates each with up to three impurity pairs; these are needed for Appendix \ref{app:MatrixElements} and Sec. \ref{sec-formalism}.
%In prior work\cite{Lei2021}, %using commutators,
%\begin{eqnarray}
%[\hat{A}_k, \hat{B}^\dagger_n] = -\frac{1}{2}trace(A_k B_n) + \hat{Q}, ~~ Q = B_n A_k,  \\
%[\hat{A}_i, ]
%\end{eqnarray}
%
We begin with a general formula that expresses the overlap for $n$-pair states in terms of matrix traces~\cite{PhysRevC.26.2640,Ginocchio19951861},
\begin{eqnarray}
&& \langle \hat{A}_1 \cdots \hat{A}_n | \hat{B}^\dagger_1 \cdots \hat{B}^\dagger_n \rangle 
%\nonumber\\
= \sum\limits^n_{k=1} (-2)^{-k} \sum\limits^{m_1 + \cdots + m_k = n}_{m_1 \geq m_2 \geq \cdots \geq m_k}
\prod_{w=1}^k m_w^{-1} \prod_{w \in \{m_1, \cdots m_k\}} (R_w !)^{-1}
\nonumber\\
&&
\times
\sum\limits_{\left\{ j_1, \cdots, j_n \right\}, \left\{i_1, \cdots, i_n\right\} } 
  \mathrm{tr} \, ( A_{i_1} B_{j_1} \cdots A_{i_{m_1}} B_{j_{m_1}} )
 \cdots
 \mathrm{tr} \,( A_{i_{n+1-m_k}} B_{j_{n+1-m_k}} \cdots A_{i_n} B_{j_n} ).
 \label{eqn:ove-explicit}
\end{eqnarray}
Here \(\{i_1, \cdots, i_n \}\) and \(\{ j_1, \cdots, j_n \}\) are rearrangements of \(\left\{1,2,\cdots,n \right\}\), $m_1 \geq m_2 \geq \cdots \geq m_k$ is a partition of $n$, and $R_i$ is the degeneracy of $m_i$ in a partition, i.e., the number of times $m_i$ is repeated.

%
%that expresses the overlap for $n$-pair states in terms of overlaps for $(n-1)$-pair states~\cite{Lei2021,PhysRevC.26.2640},
%a recursive formula for overlap between general pair configurations is rederived (see Ref. \cite{PhysRevC.26.2640}), by invoking commutators,
\begin{widetext}

While one can carry out Eq. (\ref{eqn:ove-explicit}) either by commutation relations~\cite{Lei2021,PhysRevC.26.2640} or by a generating function~\cite{Ginocchio19951861},
here we consider the important special case of a collective pair condensate, i.e., all $n$ pairs in a state are identical:
%In this work we apply this formula to collective-pair condensates, which is a special case of (\ref{eqn:ove-npa-explicit}), i.e. all pairs in the left/right ket are identical,
\begin{eqnarray}
 \langle (\hat{A}_1)^n | ( \hat{A}_2^\dagger )^n \rangle \nonumber
&=&  (n!)^2 
\sum\limits_{k=1}^n (-2)^{-k} 
\sum\limits^{m_1 + \cdots + m_k = n}_{m_1 \geq m_2 \geq \cdots \geq m_k}
\prod_{w=1}^k m_w^{-1} \prod_{w \in \{m_1, \cdots m_k\}} (R_w !)^{-1} \cdots
\nonumber\\
&&\times \mathrm{tr}\,( (A_1 A_2)^{m_1} ) \cdots \mathrm{tr}\,((A_1 A_2)^{m_k}).
\label{eqn-PCOve}
\end{eqnarray}

Therefore, to calculate one overlap between condensates as in Eq. (\ref{eqn-PCOve}), we calculate $ \mathrm{tr}\,(A_1
A_2)$, $\cdots$, $\cdots \mathrm{tr}\,( (A_1
A_2)^n ) $, and sum over all ordered partitions $\{ m_1, \cdots, m_k \}$ . 
%If one stores intermediate terms such as $trace( (A_1 A_2)^i ) , i=1,\cdots,n$, 
The computation time is polynomial  in $n$. 
Note when $n$ is large, e.g. \(n=16\) in
$2s1d0g_{7/2}0h_{11/2}$ shell (N,Z $\in [50,82]$),  the summation of (\ref{eqn-PCOve}) can have
cancellations of large numbers, requiring careful consideration of round-off error.

%Similarly, following Eq. (\ref{eqn:ove-npa-explicit}), overlap formulas with ``impurity" pairs in pair condensates can be derived.
Similarly, from Eq. (\ref{eqn:ove-explicit}), we can obtain overlap formulas with impurity pairs.
%The more impurity pairs  the more complicated the overlap formula is, when expanded.
%However, according to analysis in Appendix \ref{app:MatrixElements}, 
Our work here requires overlap formulas with at most three different impurity pairs:
%We enlist them below, as they are used in our codes.
\begin{eqnarray}
&&\langle (\hat{A}_1)^{n-1} A_3 | ( \hat{A}_2^\dagger )^{n} \rangle
\nonumber\\
&=& (n-1)! n!
\sum\limits_{k=1}^N (-2)^{-k} 
\sum\limits^{m_1 + \cdots + m_k = n}_{m_1 \geq m_2 \geq \cdots \geq m_k}
\prod_{w=1}^k m_w^{-1} \prod_{w \in \{m_1, \cdots m_k\}} (R_w !)^{-1}
\sum\limits^k_{r=1} m_r  
\nonumber\\
&& \times
\mathrm{tr}\,((A_1 A_2)^{m_1} ) \cdots \mathrm{tr}\,( (A_1 A_2)^{m_r -1} A_3 A_2 ) \cdots \mathrm{tr}\,( (A_1 A_2)^{m_k} ).
\label{eqn-PCOveMix10}
\end{eqnarray}
    \begin{eqnarray}
&& \langle (\hat{A}_1)^{n-1} A_3 | A^\dagger_4 ( \hat{A}_2^\dagger )^{n-1} \rangle
\nonumber\\
&=& ((n-1)!)^2 
\sum\limits_{k=1}^n (-2)^{-k} 
\sum\limits^{m_1 + \cdots + m_k = n}_{m_1 \geq m_2 \geq \cdots \geq m_k}
\prod_{w=1}^k m_w^{-1} \prod_{w \in \{m_1, \cdots m_k\}} (R_w !)^{-1}
\nonumber\\
&& \times ~~ \left\{ \sum\limits^k_{r=1} m_r 
\sum\limits^{m_r -1}_{s=0} \cdots \mathrm{tr}\,( A_3 (A_2 A_1)^s A_4 (A_1 A_2)^{m_r -1 -s} ) \cdots
\right.
\nonumber\\
&& \left.
+ \sum\limits^{r \neq s}_{1\leq r,s \leq k} m_r m_s \cdots \mathrm{tr}\,( (A_1 A_2)^{m_r-1}A_3 A_2 ) \cdots \mathrm{tr}\,( A_1 A_4 (A_1 A_2)^{m_s -1} ) \cdots \right\}.
\label{eqn-PCOveMix11}
\end{eqnarray}
\begin{eqnarray}
&& \langle (\hat{A}_1)^{n-2} A_3 A_4 | ( \hat{A}_2^\dagger )^{n} \rangle
\nonumber\\
&=& n! (n-2)! 
\sum\limits_{k=1}^n (-2)^{-k} 
\sum\limits^{m_1 + \cdots + m_k = n}_{m_1 \geq m_2 \geq \cdots \geq m_k}
\prod_{w=1}^k m_w^{-1} \prod_{w \in \{m_1, \cdots m_k\}} (R_w !)^{-1}
\nonumber\\
&& \times ~~ \left\{ \sum\limits^k_{r=1} m_r 
\sum\limits^{m_r -2}_{s=0} \cdots \mathrm{tr}\,( A_3 A_2 (A_1 A_2)^s A_4 A_2 (A_1 A_2)^{m_r -2 -s} ) \cdots
\right.
\nonumber\\
&& \left.
+ \sum\limits^{r \neq s}_{1\leq r,s \leq k} m_r m_s \cdots \mathrm{tr}\,( (A_1 A_2)^{m_r-1}A_3 A_2 ) \cdots \mathrm{tr}\,( A_4 A_2 (A_1 A_2)^{m_r -1} ) \cdots \right\}.
\label{eqn-PCOveMix20}
\end{eqnarray}
\begin{eqnarray}
&& \langle (\hat{A}_1)^{n-2} \hat{A}_3 \hat{A}_4 | \hat{A}^\dagger_5 ( \hat{A}_2^\dagger )^{n-1} \rangle
\nonumber\\
&=& (n-1)! (n-2)! 
\sum\limits_{k=1}^n (-2)^{-k} 
\sum\limits^{m_1 + \cdots + m_k = n}_{m_1 \geq m_2 \geq \cdots \geq m_k}
\prod_{w=1}^k m_w^{-1} \prod_{w \in \{m_1, \cdots m_k\}} (R_w !)^{-1}
\nonumber\\
&&
\left\{ 
\sum\limits^{ {r \neq s \neq t} }_{ 1\leq r,s,t \leq k} m_r m_s m_t
 \cdots \mathrm{tr}\,( A_3 A_2 (A_1 A_2)^{m_r -1} ) \cdots \mathrm{tr}\,( A_4 A_2 (A_1 A_2)^{m_s -1} ) \cdots \mathrm{tr}\,( A_1 A_5 (A_1 A_2)^{m_t -1} ) \cdots
\right.
\nonumber\\
&& +
\sum\limits^{ r \neq s }_{1 \leq r,s \leq k} m_r m_s \sum\limits^{ m_r -1 }_{ t=0 } 
 \cdots \mathrm{tr}\,( A_3 (A_2 A_1)^t A_5 (A_1 A_2)^{m_r -1 -t} ) \cdots \mathrm{tr}\,( A_4 A_2 (A_1 A_2)^{m_s -1} ) \cdots 
\nonumber\\
&& +
 \sum\limits^{ r \neq s }_{1 \leq r,s \leq k} m_r m_s \sum\limits^{ m_r -1 }_{ t=0 } 
 \cdots \mathrm{tr}\,( A_4 (A_2 A_1)^t A_5 (A_1 A_2)^{m_r -1 -t} ) \cdots \mathrm{tr}\,( A_3 A_2 (A_1 A_2)^{m_s -1} ) \cdots 
\nonumber\\
&& +
 \sum\limits^{ r \neq s }_{1 \leq r,s \leq k} m_r m_s \sum\limits^{ m_r -2 }_{ t=0 } 
 \cdots \mathrm{tr}\,( A_3 A_2 (A_1 A_2)^t A_4 A_2 (A_1 A_2)^{m_r -2 -t} ) \cdots \mathrm{tr}\,( A_1 A_5 (A_1 A_2)^{m_s -1} ) \cdots 
\nonumber\\
&& \left. +
\sum\limits^{k}_{r=1} m_r \sum\limits^{m_r -1}_{s=1} \sum\limits^{m_r}_{t=1}
\cdots \mathrm{tr}\,( A_3 A_2 A_1 A_2 \cdots [A_5]_{2t} \cdots [A_4]_{2s+1} \cdots A_1 A_2 ) \cdots
\right\},
\label{eqn-PCOveMix21}
\end{eqnarray}
where \([A_5]_{2t}\) means \(A_5\) is at position \(2t\),
\([A_4]_{2s+1}\) means $A_4$ is at position \(2s+1\) in the matrix product.
%These formulas are used in computation of Hamiltonian matrix elements and their derivatives, in variation and projection, therefore they form cornerstones of this work.

\end{widetext}

\bibliographystyle{apsrev4-1}
\bibliography{ref-papers}

\begin{thebibliography}{64}%
\makeatletter
\providecommand \@ifxundefined [1]{%
 \@ifx{#1\undefined}
}%
\providecommand \@ifnum [1]{%
 \ifnum #1\expandafter \@firstoftwo
 \else \expandafter \@secondoftwo
 \fi
}%
\providecommand \@ifx [1]{%
 \ifx #1\expandafter \@firstoftwo
 \else \expandafter \@secondoftwo
 \fi
}%
\providecommand \natexlab [1]{#1}%
\providecommand \enquote  [1]{``#1''}%
\providecommand \bibnamefont  [1]{#1}%
\providecommand \bibfnamefont [1]{#1}%
\providecommand \citenamefont [1]{#1}%
\providecommand \href@noop [0]{\@secondoftwo}%
\providecommand \href [0]{\begingroup \@sanitize@url \@href}%
\providecommand \@href[1]{\@@startlink{#1}\@@href}%
\providecommand \@@href[1]{\endgroup#1\@@endlink}%
\providecommand \@sanitize@url [0]{\catcode `\\12\catcode `\$12\catcode
  `\&12\catcode `\#12\catcode `\^12\catcode `\_12\catcode `\%12\relax}%
\providecommand \@@startlink[1]{}%
\providecommand \@@endlink[0]{}%
\providecommand \url  [0]{\begingroup\@sanitize@url \@url }%
\providecommand \@url [1]{\endgroup\@href {#1}{\urlprefix }}%
\providecommand \urlprefix  [0]{URL }%
\providecommand \Eprint [0]{\href }%
\providecommand \doibase [0]{http://dx.doi.org/}%
\providecommand \selectlanguage [0]{\@gobble}%
\providecommand \bibinfo  [0]{\@secondoftwo}%
\providecommand \bibfield  [0]{\@secondoftwo}%
\providecommand \translation [1]{[#1]}%
\providecommand \BibitemOpen [0]{}%
\providecommand \bibitemStop [0]{}%
\providecommand \bibitemNoStop [0]{.\EOS\space}%
\providecommand \EOS [0]{\spacefactor3000\relax}%
\providecommand \BibitemShut  [1]{\csname bibitem#1\endcsname}%
\let\auto@bib@innerbib\@empty
%</preamble>
\bibitem [{\citenamefont {Bohr}\ and\ \citenamefont
  {Mottelson}(1998)}]{bohr1998nuclear}%
  \BibitemOpen
  \bibfield  {author} {\bibinfo {author} {\bibfnamefont {A.~N.}\ \bibnamefont
  {Bohr}}\ and\ \bibinfo {author} {\bibfnamefont {B.~R.}\ \bibnamefont
  {Mottelson}},\ }\href@noop {} {\emph {\bibinfo {title} {Nuclear Structure (In
  2 Volumes)}}}\ (\bibinfo  {publisher} {World Scientific Publishing Company,
  Singapore},\ \bibinfo {year} {1998})\BibitemShut {NoStop}%
\bibitem [{\citenamefont {Ring}\ and\ \citenamefont {Shuck}(1980)}]{Ring1980}%
  \BibitemOpen
  \bibfield  {author} {\bibinfo {author} {\bibfnamefont {P.}~\bibnamefont
  {Ring}}\ and\ \bibinfo {author} {\bibfnamefont {P.}~\bibnamefont {Shuck}},\
  }\href@noop {} {\emph {\bibinfo {title} {The Nuclear Many-Body Problem}}}\
  (\bibinfo  {publisher} {Springe},\ \bibinfo {year} {1980})\BibitemShut
  {NoStop}%
\bibitem [{\citenamefont {Frauendorf}\ and\ \citenamefont
  {D\"onau}(2014)}]{PhysRevC.89.014322}%
  \BibitemOpen
  \bibfield  {author} {\bibinfo {author} {\bibfnamefont {S.}~\bibnamefont
  {Frauendorf}}\ and\ \bibinfo {author} {\bibfnamefont {F.}~\bibnamefont
  {D\"onau}},\ }\href {\doibase 10.1103/PhysRevC.89.014322} {\bibfield
  {journal} {\bibinfo  {journal} {Phys. Rev. C}\ }\textbf {\bibinfo {volume}
  {89}},\ \bibinfo {pages} {014322} (\bibinfo {year} {2014})}\BibitemShut
  {NoStop}%
\bibitem [{\citenamefont {Ayangeakaa}\ \emph {et~al.}(2013)\citenamefont
  {Ayangeakaa}, \citenamefont {Garg}, \citenamefont {Anthony}, \citenamefont
  {Frauendorf}, \citenamefont {Matta}, \citenamefont {Nayak}, \citenamefont
  {Patel}, \citenamefont {Chen}, \citenamefont {Zhang}, \citenamefont {Zhao},
  \citenamefont {Qi}, \citenamefont {Meng}, \citenamefont {Janssens},
  \citenamefont {Carpenter}, \citenamefont {Chiara}, \citenamefont {Kondev},
  \citenamefont {Lauritsen}, \citenamefont {Seweryniak}, \citenamefont {Zhu},
  \citenamefont {Ghugre},\ and\ \citenamefont
  {Palit}}]{PhysRevLett.110.172504}%
  \BibitemOpen
  \bibfield  {author} {\bibinfo {author} {\bibfnamefont {A.~D.}\ \bibnamefont
  {Ayangeakaa}}, \bibinfo {author} {\bibfnamefont {U.}~\bibnamefont {Garg}},
  \bibinfo {author} {\bibfnamefont {M.~D.}\ \bibnamefont {Anthony}}, \bibinfo
  {author} {\bibfnamefont {S.}~\bibnamefont {Frauendorf}}, \bibinfo {author}
  {\bibfnamefont {J.~T.}\ \bibnamefont {Matta}}, \bibinfo {author}
  {\bibfnamefont {B.~K.}\ \bibnamefont {Nayak}}, \bibinfo {author}
  {\bibfnamefont {D.}~\bibnamefont {Patel}}, \bibinfo {author} {\bibfnamefont
  {Q.~B.}\ \bibnamefont {Chen}}, \bibinfo {author} {\bibfnamefont {S.~Q.}\
  \bibnamefont {Zhang}}, \bibinfo {author} {\bibfnamefont {P.~W.}\ \bibnamefont
  {Zhao}}, \bibinfo {author} {\bibfnamefont {B.}~\bibnamefont {Qi}}, \bibinfo
  {author} {\bibfnamefont {J.}~\bibnamefont {Meng}}, \bibinfo {author}
  {\bibfnamefont {R.~V.~F.}\ \bibnamefont {Janssens}}, \bibinfo {author}
  {\bibfnamefont {M.~P.}\ \bibnamefont {Carpenter}}, \bibinfo {author}
  {\bibfnamefont {C.~J.}\ \bibnamefont {Chiara}}, \bibinfo {author}
  {\bibfnamefont {F.~G.}\ \bibnamefont {Kondev}}, \bibinfo {author}
  {\bibfnamefont {T.}~\bibnamefont {Lauritsen}}, \bibinfo {author}
  {\bibfnamefont {D.}~\bibnamefont {Seweryniak}}, \bibinfo {author}
  {\bibfnamefont {S.}~\bibnamefont {Zhu}}, \bibinfo {author} {\bibfnamefont
  {S.~S.}\ \bibnamefont {Ghugre}}, \ and\ \bibinfo {author} {\bibfnamefont
  {R.}~\bibnamefont {Palit}},\ }\href {\doibase 10.1103/PhysRevLett.110.172504}
  {\bibfield  {journal} {\bibinfo  {journal} {Phys. Rev. Lett.}\ }\textbf
  {\bibinfo {volume} {110}},\ \bibinfo {pages} {172504} (\bibinfo {year}
  {2013})}\BibitemShut {NoStop}%
\bibitem [{\citenamefont {Xu}\ \emph {et~al.}(2017)\citenamefont {Xu},
  \citenamefont {R\"opke}, \citenamefont {Schuck}, \citenamefont {Ren},
  \citenamefont {Funaki}, \citenamefont {Horiuchi}, \citenamefont {Tohsaki},
  \citenamefont {Yamada},\ and\ \citenamefont {Zhou}}]{PhysRevC.95.061306}%
  \BibitemOpen
  \bibfield  {author} {\bibinfo {author} {\bibfnamefont {C.}~\bibnamefont
  {Xu}}, \bibinfo {author} {\bibfnamefont {G.}~\bibnamefont {R\"opke}},
  \bibinfo {author} {\bibfnamefont {P.}~\bibnamefont {Schuck}}, \bibinfo
  {author} {\bibfnamefont {Z.}~\bibnamefont {Ren}}, \bibinfo {author}
  {\bibfnamefont {Y.}~\bibnamefont {Funaki}}, \bibinfo {author} {\bibfnamefont
  {H.}~\bibnamefont {Horiuchi}}, \bibinfo {author} {\bibfnamefont
  {A.}~\bibnamefont {Tohsaki}}, \bibinfo {author} {\bibfnamefont
  {T.}~\bibnamefont {Yamada}}, \ and\ \bibinfo {author} {\bibfnamefont
  {B.}~\bibnamefont {Zhou}},\ }\href {\doibase 10.1103/PhysRevC.95.061306}
  {\bibfield  {journal} {\bibinfo  {journal} {Phys. Rev. C}\ }\textbf {\bibinfo
  {volume} {95}},\ \bibinfo {pages} {061306(R)} (\bibinfo {year}
  {2017})}\BibitemShut {NoStop}%
\bibitem [{\citenamefont {Surman}\ \emph {et~al.}(2016)\citenamefont {Surman},
  \citenamefont {R.}, \citenamefont {Mumpower}, \citenamefont {M.},
  \citenamefont {R.}, \citenamefont {Aprahamian}, \citenamefont {A.},
  \citenamefont {McLaughlin}, \citenamefont {G.},\ and\ \citenamefont
  {C.}}]{Surman2016The}%
  \BibitemOpen
  \bibfield  {author} {\bibinfo {author} {\bibnamefont {Surman}}, \bibinfo
  {author} {\bibnamefont {R.}}, \bibinfo {author} {\bibnamefont {Mumpower}},
  \bibinfo {author} {\bibnamefont {M.}}, \bibinfo {author} {\bibnamefont {R.}},
  \bibinfo {author} {\bibnamefont {Aprahamian}}, \bibinfo {author}
  {\bibnamefont {A.}}, \bibinfo {author} {\bibnamefont {McLaughlin}}, \bibinfo
  {author} {\bibnamefont {G.}}, \ and\ \bibinfo {author} {\bibnamefont {C.}},\
  }\href@noop {} {\bibfield  {journal} {\bibinfo  {journal} {Progress in
  Particle \& Nuclear Physics}\ } (\bibinfo {year} {2016})}\BibitemShut
  {NoStop}%
\bibitem [{\citenamefont {Kajino}\ \emph {et~al.}(2019)\citenamefont {Kajino},
  \citenamefont {Aoki}, \citenamefont {Balantekin}, \citenamefont {Diehl},
  \citenamefont {Famiano},\ and\ \citenamefont {Mathews}}]{KAJINO2019109}%
  \BibitemOpen
  \bibfield  {author} {\bibinfo {author} {\bibfnamefont {T.}~\bibnamefont
  {Kajino}}, \bibinfo {author} {\bibfnamefont {W.}~\bibnamefont {Aoki}},
  \bibinfo {author} {\bibfnamefont {A.}~\bibnamefont {Balantekin}}, \bibinfo
  {author} {\bibfnamefont {R.}~\bibnamefont {Diehl}}, \bibinfo {author}
  {\bibfnamefont {M.}~\bibnamefont {Famiano}}, \ and\ \bibinfo {author}
  {\bibfnamefont {G.}~\bibnamefont {Mathews}},\ }\href {\doibase
  https://doi.org/10.1016/j.ppnp.2019.02.008} {\bibfield  {journal} {\bibinfo
  {journal} {Progress in Particle and Nuclear Physics}\ }\textbf {\bibinfo
  {volume} {107}},\ \bibinfo {pages} {109 } (\bibinfo {year}
  {2019})}\BibitemShut {NoStop}%
\bibitem [{\citenamefont {Meng}(2016)}]{2016Relativistic}%
  \BibitemOpen
  \bibfield  {author} {\bibinfo {author} {\bibfnamefont {J.}~\bibnamefont
  {Meng}},\ }\href@noop {} {\emph {\bibinfo {title} {International Review of
  Nuclear Physics-Vol. 10: Relativistic Density Functional for Nuclear
  Structure}}}\ (\bibinfo  {publisher} {World Scientific},\ \bibinfo {year}
  {2016})\BibitemShut {NoStop}%
\bibitem [{\citenamefont {Bally}\ and\ \citenamefont
  {Bender}(2021)}]{PhysRevC.103.024315}%
  \BibitemOpen
  \bibfield  {author} {\bibinfo {author} {\bibfnamefont {B.}~\bibnamefont
  {Bally}}\ and\ \bibinfo {author} {\bibfnamefont {M.}~\bibnamefont {Bender}},\
  }\href {\doibase 10.1103/PhysRevC.103.024315} {\bibfield  {journal} {\bibinfo
   {journal} {Phys. Rev. C}\ }\textbf {\bibinfo {volume} {103}},\ \bibinfo
  {pages} {024315} (\bibinfo {year} {2021})}\BibitemShut {NoStop}%
\bibitem [{\citenamefont {S\'anchez-Fern\'andez}\ \emph
  {et~al.}(2021)\citenamefont {S\'anchez-Fern\'andez}, \citenamefont {Bally},\
  and\ \citenamefont {Rodr\'{\i}guez}}]{PhysRevC.104.054306}%
  \BibitemOpen
  \bibfield  {author} {\bibinfo {author} {\bibfnamefont {A.}~\bibnamefont
  {S\'anchez-Fern\'andez}}, \bibinfo {author} {\bibfnamefont {B.}~\bibnamefont
  {Bally}}, \ and\ \bibinfo {author} {\bibfnamefont {T.~R.}\ \bibnamefont
  {Rodr\'{\i}guez}},\ }\href {\doibase 10.1103/PhysRevC.104.054306} {\bibfield
  {journal} {\bibinfo  {journal} {Phys. Rev. C}\ }\textbf {\bibinfo {volume}
  {104}},\ \bibinfo {pages} {054306} (\bibinfo {year} {2021})}\BibitemShut
  {NoStop}%
\bibitem [{\citenamefont {Bally}\ \emph {et~al.}(2019)\citenamefont {Bally},
  \citenamefont {S\'anchez-Fern\'andez},\ and\ \citenamefont
  {Rodr\'{\i}guez}}]{PhysRevC.100.044308}%
  \BibitemOpen
  \bibfield  {author} {\bibinfo {author} {\bibfnamefont {B.}~\bibnamefont
  {Bally}}, \bibinfo {author} {\bibfnamefont {A.}~\bibnamefont
  {S\'anchez-Fern\'andez}}, \ and\ \bibinfo {author} {\bibfnamefont {T.~R.}\
  \bibnamefont {Rodr\'{\i}guez}},\ }\href {\doibase
  10.1103/PhysRevC.100.044308} {\bibfield  {journal} {\bibinfo  {journal}
  {Phys. Rev. C}\ }\textbf {\bibinfo {volume} {100}},\ \bibinfo {pages}
  {044308} (\bibinfo {year} {2019})}\BibitemShut {NoStop}%
\bibitem [{\citenamefont {Shimizu}\ \emph {et~al.}(2021)\citenamefont
  {Shimizu}, \citenamefont {Mizusaki}, \citenamefont {Kaneko},\ and\
  \citenamefont {Tsunoda}}]{PhysRevC.103.064302}%
  \BibitemOpen
  \bibfield  {author} {\bibinfo {author} {\bibfnamefont {N.}~\bibnamefont
  {Shimizu}}, \bibinfo {author} {\bibfnamefont {T.}~\bibnamefont {Mizusaki}},
  \bibinfo {author} {\bibfnamefont {K.}~\bibnamefont {Kaneko}}, \ and\ \bibinfo
  {author} {\bibfnamefont {Y.}~\bibnamefont {Tsunoda}},\ }\href {\doibase
  10.1103/PhysRevC.103.064302} {\bibfield  {journal} {\bibinfo  {journal}
  {Phys. Rev. C}\ }\textbf {\bibinfo {volume} {103}},\ \bibinfo {pages}
  {064302} (\bibinfo {year} {2021})}\BibitemShut {NoStop}%
\bibitem [{\citenamefont {Iachello}\ and\ \citenamefont
  {Arima}(2006)}]{Iachello2006}%
  \BibitemOpen
  \bibfield  {author} {\bibinfo {author} {\bibfnamefont {F.}~\bibnamefont
  {Iachello}}\ and\ \bibinfo {author} {\bibfnamefont {A.}~\bibnamefont
  {Arima}},\ }\href@noop {} {\emph {\bibinfo {title} {The Interacting Boson
  Model}}}\ (\bibinfo  {publisher} {Cambridge Univ Pr},\ \bibinfo {year}
  {2006})\BibitemShut {NoStop}%
\bibitem [{\citenamefont {Sun}\ \emph {et~al.}(2000)\citenamefont {Sun},
  \citenamefont {Hara}, \citenamefont {Sheikh}, \citenamefont {Hirsch},
  \citenamefont {Vel\'azquez},\ and\ \citenamefont
  {Guidry}}]{PhysRevC.61.064323}%
  \BibitemOpen
  \bibfield  {author} {\bibinfo {author} {\bibfnamefont {Y.}~\bibnamefont
  {Sun}}, \bibinfo {author} {\bibfnamefont {K.}~\bibnamefont {Hara}}, \bibinfo
  {author} {\bibfnamefont {J.~A.}\ \bibnamefont {Sheikh}}, \bibinfo {author}
  {\bibfnamefont {J.~G.}\ \bibnamefont {Hirsch}}, \bibinfo {author}
  {\bibfnamefont {V.}~\bibnamefont {Vel\'azquez}}, \ and\ \bibinfo {author}
  {\bibfnamefont {M.}~\bibnamefont {Guidry}},\ }\href {\doibase
  10.1103/PhysRevC.61.064323} {\bibfield  {journal} {\bibinfo  {journal} {Phys.
  Rev. C}\ }\textbf {\bibinfo {volume} {61}},\ \bibinfo {pages} {064323}
  (\bibinfo {year} {2000})}\BibitemShut {NoStop}%
\bibitem [{\citenamefont {Otsuka}\ \emph {et~al.}(2001)\citenamefont {Otsuka},
  \citenamefont {Honma}, \citenamefont {Mizusaki}, \citenamefont {Shimizu},\
  and\ \citenamefont {Utsuno}}]{mcsm-pair}%
  \BibitemOpen
  \bibfield  {author} {\bibinfo {author} {\bibfnamefont {T.}~\bibnamefont
  {Otsuka}}, \bibinfo {author} {\bibfnamefont {M.}~\bibnamefont {Honma}},
  \bibinfo {author} {\bibfnamefont {T.}~\bibnamefont {Mizusaki}}, \bibinfo
  {author} {\bibfnamefont {N.}~\bibnamefont {Shimizu}}, \ and\ \bibinfo
  {author} {\bibfnamefont {Y.}~\bibnamefont {Utsuno}},\ }\href {\doibase
  https://doi.org/10.1016/S0146-6410(01)00157-0} {\bibfield  {journal}
  {\bibinfo  {journal} {Prog. Part. Nucl. Phys.}\ }\textbf {\bibinfo {volume}
  {47}},\ \bibinfo {pages} {319 } (\bibinfo {year} {2001})}\BibitemShut
  {NoStop}%
\bibitem [{\citenamefont {Caurier}\ \emph {et~al.}(2005)\citenamefont
  {Caurier}, \citenamefont {Mart\'{\i}nez-Pinedo}, \citenamefont {Nowacki},
  \citenamefont {Poves},\ and\ \citenamefont {Zuker}}]{RevModPhys.77.427}%
  \BibitemOpen
  \bibfield  {author} {\bibinfo {author} {\bibfnamefont {E.}~\bibnamefont
  {Caurier}}, \bibinfo {author} {\bibfnamefont {G.}~\bibnamefont
  {Mart\'{\i}nez-Pinedo}}, \bibinfo {author} {\bibfnamefont {F.}~\bibnamefont
  {Nowacki}}, \bibinfo {author} {\bibfnamefont {A.}~\bibnamefont {Poves}}, \
  and\ \bibinfo {author} {\bibfnamefont {A.~P.}\ \bibnamefont {Zuker}},\ }\href
  {\doibase 10.1103/RevModPhys.77.427} {\bibfield  {journal} {\bibinfo
  {journal} {Rev. Mod. Phys.}\ }\textbf {\bibinfo {volume} {77}},\ \bibinfo
  {pages} {427} (\bibinfo {year} {2005})}\BibitemShut {NoStop}%
\bibitem [{\citenamefont {Brown}(2001)}]{BROWN2001517}%
  \BibitemOpen
  \bibfield  {author} {\bibinfo {author} {\bibfnamefont {B.}~\bibnamefont
  {Brown}},\ }\href {\doibase https://doi.org/10.1016/S0146-6410(01)00159-4}
  {\bibfield  {journal} {\bibinfo  {journal} {Progress in Particle and Nuclear
  Physics}\ }\textbf {\bibinfo {volume} {47}},\ \bibinfo {pages} {517}
  (\bibinfo {year} {2001})}\BibitemShut {NoStop}%
\bibitem [{\citenamefont {Brown}\ and\ \citenamefont
  {Richter}(2006)}]{PhysRevC.74.034315}%
  \BibitemOpen
  \bibfield  {author} {\bibinfo {author} {\bibfnamefont {B.~A.}\ \bibnamefont
  {Brown}}\ and\ \bibinfo {author} {\bibfnamefont {W.~A.}\ \bibnamefont
  {Richter}},\ }\href {\doibase 10.1103/PhysRevC.74.034315} {\bibfield
  {journal} {\bibinfo  {journal} {Phys. Rev. C}\ }\textbf {\bibinfo {volume}
  {74}},\ \bibinfo {pages} {034315} (\bibinfo {year} {2006})}\BibitemShut
  {NoStop}%
\bibitem [{\citenamefont {Honma}\ \emph {et~al.}(2002)\citenamefont {Honma},
  \citenamefont {Otsuka}, \citenamefont {Brown},\ and\ \citenamefont
  {Mizusaki}}]{PhysRevC.65.061301}%
  \BibitemOpen
  \bibfield  {author} {\bibinfo {author} {\bibfnamefont {M.}~\bibnamefont
  {Honma}}, \bibinfo {author} {\bibfnamefont {T.}~\bibnamefont {Otsuka}},
  \bibinfo {author} {\bibfnamefont {B.~A.}\ \bibnamefont {Brown}}, \ and\
  \bibinfo {author} {\bibfnamefont {T.}~\bibnamefont {Mizusaki}},\ }\href
  {\doibase 10.1103/PhysRevC.65.061301} {\bibfield  {journal} {\bibinfo
  {journal} {Phys. Rev. C}\ }\textbf {\bibinfo {volume} {65}},\ \bibinfo
  {pages} {061301(R)} (\bibinfo {year} {2002})}\BibitemShut {NoStop}%
\bibitem [{\citenamefont {Honma}\ \emph {et~al.}(2005)\citenamefont {Honma},
  \citenamefont {Otsuka}, \citenamefont {Brown},\ and\ \citenamefont
  {Mizusaki}}]{Honma2005}%
  \BibitemOpen
  \bibfield  {author} {\bibinfo {author} {\bibfnamefont {M.}~\bibnamefont
  {Honma}}, \bibinfo {author} {\bibfnamefont {T.}~\bibnamefont {Otsuka}},
  \bibinfo {author} {\bibfnamefont {B.~A.}\ \bibnamefont {Brown}}, \ and\
  \bibinfo {author} {\bibfnamefont {T.}~\bibnamefont {Mizusaki}},\ }\href@noop
  {} {\bibfield  {journal} {\bibinfo  {journal} {The European Physical Journal
  A - Hadrons and Nuclei}\ }\textbf {\bibinfo {volume} {25}},\ \bibinfo {pages}
  {499} (\bibinfo {year} {2005})}\BibitemShut {NoStop}%
\bibitem [{\citenamefont {Talmi}(1993)}]{1981Simple}%
  \BibitemOpen
  \bibfield  {author} {\bibinfo {author} {\bibfnamefont {I.}~\bibnamefont
  {Talmi}},\ }\href@noop {} {\emph {\bibinfo {title} {Simple Models of Complex
  Nuclei (Contemporary Concepts in Physics)}}}\ (\bibinfo  {publisher} {CRC
  Press},\ \bibinfo {year} {1993})\BibitemShut {NoStop}%
\bibitem [{\citenamefont {Jia}(2019)}]{PhysRevC.99.014302}%
  \BibitemOpen
  \bibfield  {author} {\bibinfo {author} {\bibfnamefont {L.~Y.}\ \bibnamefont
  {Jia}},\ }\href {\doibase 10.1103/PhysRevC.99.014302} {\bibfield  {journal}
  {\bibinfo  {journal} {Phys. Rev. C}\ }\textbf {\bibinfo {volume} {99}},\
  \bibinfo {pages} {014302} (\bibinfo {year} {2019})}\BibitemShut {NoStop}%
\bibitem [{\citenamefont {Jia}(2016)}]{PhysRevC.93.064307}%
  \BibitemOpen
  \bibfield  {author} {\bibinfo {author} {\bibfnamefont {L.~Y.}\ \bibnamefont
  {Jia}},\ }\href {\doibase 10.1103/PhysRevC.93.064307} {\bibfield  {journal}
  {\bibinfo  {journal} {Phys. Rev. C}\ }\textbf {\bibinfo {volume} {93}},\
  \bibinfo {pages} {064307} (\bibinfo {year} {2016})}\BibitemShut {NoStop}%
\bibitem [{\citenamefont {Jia}(2015)}]{Jia_2015}%
  \BibitemOpen
  \bibfield  {author} {\bibinfo {author} {\bibfnamefont {L.~Y.}\ \bibnamefont
  {Jia}},\ }\href {\doibase 10.1088/0954-3899/42/11/115105} {\bibfield
  {journal} {\bibinfo  {journal} {Journal of Physics G: Nuclear and Particle
  Physics}\ }\textbf {\bibinfo {volume} {42}},\ \bibinfo {pages} {115105}
  (\bibinfo {year} {2015})}\BibitemShut {NoStop}%
\bibitem [{\citenamefont {Zhao}\ and\ \citenamefont {Arima}(2014)}]{ZHAO20141}%
  \BibitemOpen
  \bibfield  {author} {\bibinfo {author} {\bibfnamefont {Y.}~\bibnamefont
  {Zhao}}\ and\ \bibinfo {author} {\bibfnamefont {A.}~\bibnamefont {Arima}},\
  }\href {\doibase https://doi.org/10.1016/j.physrep.2014.07.002} {\bibfield
  {journal} {\bibinfo  {journal} {Physics Reports}\ }\textbf {\bibinfo {volume}
  {545}},\ \bibinfo {pages} {1} (\bibinfo {year} {2014})},\ \bibinfo {note}
  {nucleon-pair approximation to the nuclear shell model}\BibitemShut {NoStop}%
\bibitem [{\citenamefont {Higashiyama}\ \emph {et~al.}(2002)\citenamefont
  {Higashiyama}, \citenamefont {Yoshinaga},\ and\ \citenamefont
  {Tanabe}}]{PhysRevC.65.054317}%
  \BibitemOpen
  \bibfield  {author} {\bibinfo {author} {\bibfnamefont {K.}~\bibnamefont
  {Higashiyama}}, \bibinfo {author} {\bibfnamefont {N.}~\bibnamefont
  {Yoshinaga}}, \ and\ \bibinfo {author} {\bibfnamefont {K.}~\bibnamefont
  {Tanabe}},\ }\href {\doibase 10.1103/PhysRevC.65.054317} {\bibfield
  {journal} {\bibinfo  {journal} {Phys. Rev. C}\ }\textbf {\bibinfo {volume}
  {65}},\ \bibinfo {pages} {054317} (\bibinfo {year} {2002})}\BibitemShut
  {NoStop}%
\bibitem [{\citenamefont {He}\ \emph {et~al.}(2020)\citenamefont {He},
  \citenamefont {Li}, \citenamefont {Luo}, \citenamefont {Zhang}, \citenamefont
  {Pan},\ and\ \citenamefont {Draayer}}]{PhysRevC.102.024304}%
  \BibitemOpen
  \bibfield  {author} {\bibinfo {author} {\bibfnamefont {B.~C.}\ \bibnamefont
  {He}}, \bibinfo {author} {\bibfnamefont {L.}~\bibnamefont {Li}}, \bibinfo
  {author} {\bibfnamefont {Y.~A.}\ \bibnamefont {Luo}}, \bibinfo {author}
  {\bibfnamefont {Y.}~\bibnamefont {Zhang}}, \bibinfo {author} {\bibfnamefont
  {F.}~\bibnamefont {Pan}}, \ and\ \bibinfo {author} {\bibfnamefont {J.~P.}\
  \bibnamefont {Draayer}},\ }\href {\doibase 10.1103/PhysRevC.102.024304}
  {\bibfield  {journal} {\bibinfo  {journal} {Phys. Rev. C}\ }\textbf {\bibinfo
  {volume} {102}},\ \bibinfo {pages} {024304} (\bibinfo {year}
  {2020})}\BibitemShut {NoStop}%
\bibitem [{\citenamefont {Lei}\ \emph {et~al.}(2021)\citenamefont {Lei},
  \citenamefont {Lu},\ and\ \citenamefont {Zhao}}]{Lei2021}%
  \BibitemOpen
  \bibfield  {author} {\bibinfo {author} {\bibfnamefont {Y.}~\bibnamefont
  {Lei}}, \bibinfo {author} {\bibfnamefont {Y.}~\bibnamefont {Lu}}, \ and\
  \bibinfo {author} {\bibfnamefont {Y.~M.}\ \bibnamefont {Zhao}},\ }\href@noop
  {} {\bibfield  {journal} {\bibinfo  {journal} {Chinese Physics C 45, 054103
  (2021)}\ } (\bibinfo {year} {2021})}\BibitemShut {NoStop}%
\bibitem [{\citenamefont {Otsuka}\ \emph {et~al.}(1978)\citenamefont {Otsuka},
  \citenamefont {Arima}, \citenamefont {Iachello},\ and\ \citenamefont
  {Talmi}}]{OTSUKA1978139}%
  \BibitemOpen
  \bibfield  {author} {\bibinfo {author} {\bibfnamefont {T.}~\bibnamefont
  {Otsuka}}, \bibinfo {author} {\bibfnamefont {A.}~\bibnamefont {Arima}},
  \bibinfo {author} {\bibfnamefont {F.}~\bibnamefont {Iachello}}, \ and\
  \bibinfo {author} {\bibfnamefont {I.}~\bibnamefont {Talmi}},\ }\href
  {\doibase https://doi.org/10.1016/0370-2693(78)90260-5} {\bibfield  {journal}
  {\bibinfo  {journal} {Physics Letters B}\ }\textbf {\bibinfo {volume} {76}},\
  \bibinfo {pages} {139 } (\bibinfo {year} {1978})}\BibitemShut {NoStop}%
\bibitem [{\citenamefont {Johnson}\ and\ \citenamefont
  {Ginocchio}(1994)}]{PhysRevC.50.R571}%
  \BibitemOpen
  \bibfield  {author} {\bibinfo {author} {\bibfnamefont {C.~W.}\ \bibnamefont
  {Johnson}}\ and\ \bibinfo {author} {\bibfnamefont {J.~N.}\ \bibnamefont
  {Ginocchio}},\ }\href {\doibase 10.1103/PhysRevC.50.R571} {\bibfield
  {journal} {\bibinfo  {journal} {Phys. Rev. C}\ }\textbf {\bibinfo {volume}
  {50}},\ \bibinfo {pages} {R571} (\bibinfo {year} {1994})}\BibitemShut
  {NoStop}%
\bibitem [{\citenamefont {Ginocchio}\ and\ \citenamefont
  {Johnson}(1995)}]{Ginocchio19951861}%
  \BibitemOpen
  \bibfield  {author} {\bibinfo {author} {\bibfnamefont {J.}~\bibnamefont
  {Ginocchio}}\ and\ \bibinfo {author} {\bibfnamefont {C.}~\bibnamefont
  {Johnson}},\ }\href {\doibase 10.1103/PhysRevC.51.1861} {\bibfield  {journal}
  {\bibinfo  {journal} {Physical Review C}\ }\textbf {\bibinfo {volume} {51}},\
  \bibinfo {pages} {1861} (\bibinfo {year} {1995})},\ \bibinfo {note} {cited By
  9}\BibitemShut {NoStop}%
\bibitem [{\citenamefont {Fu}\ and\ \citenamefont
  {Johnson}(2020)}]{FU2020135705}%
  \BibitemOpen
  \bibfield  {author} {\bibinfo {author} {\bibfnamefont {G.}~\bibnamefont
  {Fu}}\ and\ \bibinfo {author} {\bibfnamefont {C.~W.}\ \bibnamefont
  {Johnson}},\ }\href {\doibase https://doi.org/10.1016/j.physletb.2020.135705}
  {\bibfield  {journal} {\bibinfo  {journal} {Physics Letters B}\ }\textbf
  {\bibinfo {volume} {809}},\ \bibinfo {pages} {135705} (\bibinfo {year}
  {2020})}\BibitemShut {NoStop}%
\bibitem [{\citenamefont {Fu}\ and\ \citenamefont
  {Johnson}(2021)}]{PhysRevC.104.024312}%
  \BibitemOpen
  \bibfield  {author} {\bibinfo {author} {\bibfnamefont {G.~J.}\ \bibnamefont
  {Fu}}\ and\ \bibinfo {author} {\bibfnamefont {C.~W.}\ \bibnamefont
  {Johnson}},\ }\href {\doibase 10.1103/PhysRevC.104.024312} {\bibfield
  {journal} {\bibinfo  {journal} {Phys. Rev. C}\ }\textbf {\bibinfo {volume}
  {104}},\ \bibinfo {pages} {024312} (\bibinfo {year} {2021})}\BibitemShut
  {NoStop}%
\bibitem [{\citenamefont {Lei}\ \emph {et~al.}(2020)\citenamefont {Lei},
  \citenamefont {Jiang},\ and\ \citenamefont {Pittel}}]{PhysRevC.102.024310}%
  \BibitemOpen
  \bibfield  {author} {\bibinfo {author} {\bibfnamefont {Y.}~\bibnamefont
  {Lei}}, \bibinfo {author} {\bibfnamefont {H.}~\bibnamefont {Jiang}}, \ and\
  \bibinfo {author} {\bibfnamefont {S.}~\bibnamefont {Pittel}},\ }\href
  {\doibase 10.1103/PhysRevC.102.024310} {\bibfield  {journal} {\bibinfo
  {journal} {Phys. Rev. C}\ }\textbf {\bibinfo {volume} {102}},\ \bibinfo
  {pages} {024310} (\bibinfo {year} {2020})}\BibitemShut {NoStop}%
\bibitem [{\citenamefont {Mizusaki}\ and\ \citenamefont
  {Schuck}(2021)}]{PhysRevC.104.L031305}%
  \BibitemOpen
  \bibfield  {author} {\bibinfo {author} {\bibfnamefont {T.}~\bibnamefont
  {Mizusaki}}\ and\ \bibinfo {author} {\bibfnamefont {P.}~\bibnamefont
  {Schuck}},\ }\href {\doibase 10.1103/PhysRevC.104.L031305} {\bibfield
  {journal} {\bibinfo  {journal} {Phys. Rev. C}\ }\textbf {\bibinfo {volume}
  {104}},\ \bibinfo {pages} {L031305} (\bibinfo {year} {2021})}\BibitemShut
  {NoStop}%
\bibitem [{\citenamefont {Coleman}(1965)}]{Coleman1965}%
  \BibitemOpen
  \bibfield  {author} {\bibinfo {author} {\bibfnamefont {A.~J.}\ \bibnamefont
  {Coleman}},\ }\href {\doibase 10.1063/1.1704794} {\bibfield  {journal}
  {\bibinfo  {journal} {Journal of Mathematical Physics}\ }\textbf {\bibinfo
  {volume} {6}},\ \bibinfo {pages} {1425} (\bibinfo {year} {1965})},\ \Eprint
  {http://arxiv.org/abs/https://doi.org/10.1063/1.1704794}
  {https://doi.org/10.1063/1.1704794} \BibitemShut {NoStop}%
\bibitem [{\citenamefont {Rowe}\ \emph {et~al.}(1991)\citenamefont {Rowe},
  \citenamefont {Song},\ and\ \citenamefont {Chen}}]{PhysRevC.44.R598}%
  \BibitemOpen
  \bibfield  {author} {\bibinfo {author} {\bibfnamefont {D.~J.}\ \bibnamefont
  {Rowe}}, \bibinfo {author} {\bibfnamefont {T.}~\bibnamefont {Song}}, \ and\
  \bibinfo {author} {\bibfnamefont {H.}~\bibnamefont {Chen}},\ }\href {\doibase
  10.1103/PhysRevC.44.R598} {\bibfield  {journal} {\bibinfo  {journal} {Phys.
  Rev. C}\ }\textbf {\bibinfo {volume} {44}},\ \bibinfo {pages} {R598}
  (\bibinfo {year} {1991})}\BibitemShut {NoStop}%
\bibitem [{\citenamefont {Chen}\ \emph {et~al.}(1995)\citenamefont {Chen},
  \citenamefont {Song},\ and\ \citenamefont {Rowe}}]{CHEN1995181}%
  \BibitemOpen
  \bibfield  {author} {\bibinfo {author} {\bibfnamefont {H.}~\bibnamefont
  {Chen}}, \bibinfo {author} {\bibfnamefont {T.}~\bibnamefont {Song}}, \ and\
  \bibinfo {author} {\bibfnamefont {D.}~\bibnamefont {Rowe}},\ }\href {\doibase
  https://doi.org/10.1016/0375-9474(94)00472-Y} {\bibfield  {journal} {\bibinfo
   {journal} {Nuclear Physics A}\ }\textbf {\bibinfo {volume} {582}},\ \bibinfo
  {pages} {181} (\bibinfo {year} {1995})}\BibitemShut {NoStop}%
\bibitem [{\citenamefont {Johnson}\ and\ \citenamefont
  {O'Mara}(2017)}]{PhysRevC.96.064304}%
  \BibitemOpen
  \bibfield  {author} {\bibinfo {author} {\bibfnamefont {C.~W.}\ \bibnamefont
  {Johnson}}\ and\ \bibinfo {author} {\bibfnamefont {K.~D.}\ \bibnamefont
  {O'Mara}},\ }\href {\doibase 10.1103/PhysRevC.96.064304} {\bibfield
  {journal} {\bibinfo  {journal} {Phys. Rev. C}\ }\textbf {\bibinfo {volume}
  {96}},\ \bibinfo {pages} {064304} (\bibinfo {year} {2017})}\BibitemShut
  {NoStop}%
\bibitem [{\citenamefont {Johnson}\ and\ \citenamefont
  {Jiao}(2019)}]{2019Convergence}%
  \BibitemOpen
  \bibfield  {author} {\bibinfo {author} {\bibfnamefont {C.~W.}\ \bibnamefont
  {Johnson}}\ and\ \bibinfo {author} {\bibfnamefont {C.}~\bibnamefont {Jiao}},\
  }\href@noop {} {\bibfield  {journal} {\bibinfo  {journal} {Journal of Physics
  G Nuclear \& Particle Physics}\ }\textbf {\bibinfo {volume} {46}} (\bibinfo
  {year} {2019})}\BibitemShut {NoStop}%
\bibitem [{\citenamefont {Qi}\ and\ \citenamefont
  {Xu}(2012)}]{PhysRevC.86.044323}%
  \BibitemOpen
  \bibfield  {author} {\bibinfo {author} {\bibfnamefont {C.}~\bibnamefont
  {Qi}}\ and\ \bibinfo {author} {\bibfnamefont {Z.~X.}\ \bibnamefont {Xu}},\
  }\href {\doibase 10.1103/PhysRevC.86.044323} {\bibfield  {journal} {\bibinfo
  {journal} {Phys. Rev. C}\ }\textbf {\bibinfo {volume} {86}},\ \bibinfo
  {pages} {044323} (\bibinfo {year} {2012})}\BibitemShut {NoStop}%
\bibitem [{\citenamefont {Galassi}\ \emph {et~al.}(2019)\citenamefont
  {Galassi}, \citenamefont {Davies}, \citenamefont {Theiler}, \citenamefont
  {Gough}, \citenamefont {Jungman}, \citenamefont {Alken}, \citenamefont
  {Booth}, \citenamefont {Rossi},\ and\ \citenamefont {Ulerich}}]{gsl}%
  \BibitemOpen
  \bibfield  {author} {\bibinfo {author} {\bibfnamefont {M.}~\bibnamefont
  {Galassi}}, \bibinfo {author} {\bibfnamefont {J.}~\bibnamefont {Davies}},
  \bibinfo {author} {\bibfnamefont {J.}~\bibnamefont {Theiler}}, \bibinfo
  {author} {\bibfnamefont {B.}~\bibnamefont {Gough}}, \bibinfo {author}
  {\bibfnamefont {G.}~\bibnamefont {Jungman}}, \bibinfo {author} {\bibfnamefont
  {P.}~\bibnamefont {Alken}}, \bibinfo {author} {\bibfnamefont
  {M.}~\bibnamefont {Booth}}, \bibinfo {author} {\bibfnamefont
  {F.}~\bibnamefont {Rossi}}, \ and\ \bibinfo {author} {\bibfnamefont
  {R.}~\bibnamefont {Ulerich}},\ }\href@noop {} {\emph {\bibinfo {title} {GNU
  Scientific Library}}}\ (\bibinfo  {publisher} {gnu},\ \bibinfo {year}
  {2019})\BibitemShut {NoStop}%
\bibitem [{\citenamefont {Fletcher}(1984)}]{1984Practical}%
  \BibitemOpen
  \bibfield  {author} {\bibinfo {author} {\bibfnamefont {R.}~\bibnamefont
  {Fletcher}},\ }\href@noop {} {\bibfield  {journal} {\bibinfo  {journal} {SIAM
  Review}\ }\textbf {\bibinfo {volume} {26}},\ \bibinfo {pages} {143–144}
  (\bibinfo {year} {1984})}\BibitemShut {NoStop}%
\bibitem [{\citenamefont {Sun}(2016)}]{Sun2016}%
  \BibitemOpen
  \bibfield  {author} {\bibinfo {author} {\bibfnamefont {Y.}~\bibnamefont
  {Sun}},\ }\href@noop {} {\bibfield  {journal} {\bibinfo  {journal} {Phys.
  Scr. 91 (2016) 043005 (23pp)}\ } (\bibinfo {year} {2016})}\BibitemShut
  {NoStop}%
\bibitem [{\citenamefont {Yao}\ \emph {et~al.}(2010)\citenamefont {Yao},
  \citenamefont {Meng}, \citenamefont {Ring},\ and\ \citenamefont
  {Vretenar}}]{J.M.Yao2010}%
  \BibitemOpen
  \bibfield  {author} {\bibinfo {author} {\bibfnamefont {J.~M.}\ \bibnamefont
  {Yao}}, \bibinfo {author} {\bibfnamefont {J.}~\bibnamefont {Meng}}, \bibinfo
  {author} {\bibfnamefont {P.}~\bibnamefont {Ring}}, \ and\ \bibinfo {author}
  {\bibfnamefont {D.}~\bibnamefont {Vretenar}},\ }\href@noop {} {\bibfield
  {journal} {\bibinfo  {journal} {Physical Review C 81, 044311}\ } (\bibinfo
  {year} {2010})}\BibitemShut {NoStop}%
\bibitem [{\citenamefont {Rodríguez-Guzmán}\ \emph
  {et~al.}(2002)\citenamefont {Rodríguez-Guzmán}, \citenamefont {Egido},\
  and\ \citenamefont {Robledo}}]{R.Rodriguez-Guzman2002}%
  \BibitemOpen
  \bibfield  {author} {\bibinfo {author} {\bibfnamefont {R.}~\bibnamefont
  {Rodríguez-Guzmán}}, \bibinfo {author} {\bibfnamefont {J.}~\bibnamefont
  {Egido}}, \ and\ \bibinfo {author} {\bibfnamefont {L.~M.}\ \bibnamefont
  {Robledo}},\ }\href@noop {} {\bibfield  {journal} {\bibinfo  {journal}
  {Nuclear Physics A 709 (2002) 201–235}\ } (\bibinfo {year}
  {2002})}\BibitemShut {NoStop}%
\bibitem [{\citenamefont {Edmonds}(1957)}]{Edmonds1957}%
  \BibitemOpen
  \bibfield  {author} {\bibinfo {author} {\bibfnamefont {A.~R.}\ \bibnamefont
  {Edmonds}},\ }\href@noop {} {\emph {\bibinfo {title} {Angular momentum in
  quantum mechanics}}},\ edited by\ \bibinfo {editor} {\bibnamefont
  {Princeton}}\ (\bibinfo  {publisher} {Princeton University Press},\ \bibinfo
  {year} {1957})\BibitemShut {NoStop}%
\bibitem [{\citenamefont {Sheikh}\ \emph {et~al.}(2002)\citenamefont {Sheikh},
  \citenamefont {Ring}, \citenamefont {Lopes},\ and\ \citenamefont
  {Rossignoli}}]{PhysRevC.66.044318}%
  \BibitemOpen
  \bibfield  {author} {\bibinfo {author} {\bibfnamefont {J.~A.}\ \bibnamefont
  {Sheikh}}, \bibinfo {author} {\bibfnamefont {P.}~\bibnamefont {Ring}},
  \bibinfo {author} {\bibfnamefont {E.}~\bibnamefont {Lopes}}, \ and\ \bibinfo
  {author} {\bibfnamefont {R.}~\bibnamefont {Rossignoli}},\ }\href {\doibase
  10.1103/PhysRevC.66.044318} {\bibfield  {journal} {\bibinfo  {journal} {Phys.
  Rev. C}\ }\textbf {\bibinfo {volume} {66}},\ \bibinfo {pages} {044318}
  (\bibinfo {year} {2002})}\BibitemShut {NoStop}%
\bibitem [{\citenamefont {Johnson}\ \emph {et~al.}(2013)\citenamefont
  {Johnson}, \citenamefont {Ormand},\ and\ \citenamefont
  {Krastev}}]{2013Factorization}%
  \BibitemOpen
  \bibfield  {author} {\bibinfo {author} {\bibfnamefont {C.~W.}\ \bibnamefont
  {Johnson}}, \bibinfo {author} {\bibfnamefont {W.}~\bibnamefont {Ormand}}, \
  and\ \bibinfo {author} {\bibfnamefont {P.~G.}\ \bibnamefont {Krastev}},\
  }\href@noop {} {\bibfield  {journal} {\bibinfo  {journal} {Computer Physics
  Communications}\ }\textbf {\bibinfo {volume} {184}} (\bibinfo {year}
  {2013})}\BibitemShut {NoStop}%
\bibitem [{\citenamefont {Lauber}\ \emph {et~al.}(2021)\citenamefont {Lauber},
  \citenamefont {Frye},\ and\ \citenamefont
  {Johnson}}]{lauber2021benchmarking}%
  \BibitemOpen
  \bibfield  {author} {\bibinfo {author} {\bibfnamefont {S.~M.}\ \bibnamefont
  {Lauber}}, \bibinfo {author} {\bibfnamefont {H.~C.}\ \bibnamefont {Frye}}, \
  and\ \bibinfo {author} {\bibfnamefont {C.~W.}\ \bibnamefont {Johnson}},\
  }\href@noop {} {\bibfield  {journal} {\bibinfo  {journal} {Journal of Physics
  G: Nuclear and Particle Physics}\ }\textbf {\bibinfo {volume} {48}},\
  \bibinfo {pages} {095107} (\bibinfo {year} {2021})}\BibitemShut {NoStop}%
\bibitem [{\citenamefont {Johnson}\ \emph {et~al.}(2020)\citenamefont
  {Johnson}, \citenamefont {Luu},\ and\ \citenamefont {Lu}}]{Johnson_2020}%
  \BibitemOpen
  \bibfield  {author} {\bibinfo {author} {\bibfnamefont {C.~W.}\ \bibnamefont
  {Johnson}}, \bibinfo {author} {\bibfnamefont {K.~A.}\ \bibnamefont {Luu}}, \
  and\ \bibinfo {author} {\bibfnamefont {Y.}~\bibnamefont {Lu}},\ }\href
  {\doibase 10.1088/1361-6471/abacda} {\bibfield  {journal} {\bibinfo
  {journal} {Journal of Physics G: Nuclear and Particle Physics}\ }\textbf
  {\bibinfo {volume} {47}},\ \bibinfo {pages} {105107} (\bibinfo {year}
  {2020})}\BibitemShut {NoStop}%
\bibitem [{\citenamefont {Nica}\ and\ \citenamefont
  {Singh}(2012)}]{NICA20121563}%
  \BibitemOpen
  \bibfield  {author} {\bibinfo {author} {\bibfnamefont {N.}~\bibnamefont
  {Nica}}\ and\ \bibinfo {author} {\bibfnamefont {B.}~\bibnamefont {Singh}},\
  }\href {\doibase https://doi.org/10.1016/j.nds.2012.06.001} {\bibfield
  {journal} {\bibinfo  {journal} {Nuclear Data Sheets}\ }\textbf {\bibinfo
  {volume} {113}},\ \bibinfo {pages} {1563} (\bibinfo {year}
  {2012})}\BibitemShut {NoStop}%
\bibitem [{\citenamefont {{Kanno}}\ \emph {et~al.}(2002)\citenamefont
  {{Kanno}}, \citenamefont {{Gomi}}, \citenamefont {{Motobayashi}},
  \citenamefont {{Yoneda}}, \citenamefont {{Aoi}}, \citenamefont {{Ando}},
  \citenamefont {{Baba}}, \citenamefont {{Demichi}}, \citenamefont {{Fulop}},
  \citenamefont {{Futakami}}, \citenamefont {{Hasegawa}}, \citenamefont
  {{Higurashi}}, \citenamefont {{Ieki}}, \citenamefont {{Imai}}, \citenamefont
  {{Iwasa}}, \citenamefont {{Iwasaki}}, \citenamefont {{Kubo}}, \citenamefont
  {{Kubono}}, \citenamefont {{Kunibu}}, \citenamefont {{Matsuyama}},
  \citenamefont {{Michimasa}}, \citenamefont {{Minemura}}, \citenamefont
  {{Murakami}}, \citenamefont {{Nakamura}}, \citenamefont {{Saito}},
  \citenamefont {{Sakurai}}, \citenamefont {{Serata}}, \citenamefont
  {{Shimoura}}, \citenamefont {{Sugimoto}}, \citenamefont {{Takeshita}},
  \citenamefont {{Takeuchi}}, \citenamefont {{Ue}}, \citenamefont {{Yamada}},
  \citenamefont {{Yanagisawa}}, \citenamefont {{Yoshida}},\ and\ \citenamefont
  {{Ishihara}}}]{NSR2002KA80}%
  \BibitemOpen
  \bibfield  {author} {\bibinfo {author} {\bibfnamefont {S.}~\bibnamefont
  {{Kanno}}}, \bibinfo {author} {\bibfnamefont {T.}~\bibnamefont {{Gomi}}},
  \bibinfo {author} {\bibfnamefont {T.}~\bibnamefont {{Motobayashi}}}, \bibinfo
  {author} {\bibfnamefont {K.}~\bibnamefont {{Yoneda}}}, \bibinfo {author}
  {\bibfnamefont {N.}~\bibnamefont {{Aoi}}}, \bibinfo {author} {\bibfnamefont
  {Y.}~\bibnamefont {{Ando}}}, \bibinfo {author} {\bibfnamefont
  {H.}~\bibnamefont {{Baba}}}, \bibinfo {author} {\bibfnamefont
  {K.}~\bibnamefont {{Demichi}}}, \bibinfo {author} {\bibfnamefont
  {Z.}~\bibnamefont {{Fulop}}}, \bibinfo {author} {\bibfnamefont
  {U.}~\bibnamefont {{Futakami}}}, \bibinfo {author} {\bibfnamefont
  {H.}~\bibnamefont {{Hasegawa}}}, \bibinfo {author} {\bibfnamefont
  {Y.}~\bibnamefont {{Higurashi}}}, \bibinfo {author} {\bibfnamefont
  {K.}~\bibnamefont {{Ieki}}}, \bibinfo {author} {\bibfnamefont
  {N.}~\bibnamefont {{Imai}}}, \bibinfo {author} {\bibfnamefont
  {N.}~\bibnamefont {{Iwasa}}}, \bibinfo {author} {\bibfnamefont
  {H.}~\bibnamefont {{Iwasaki}}}, \bibinfo {author} {\bibfnamefont
  {T.}~\bibnamefont {{Kubo}}}, \bibinfo {author} {\bibfnamefont
  {S.}~\bibnamefont {{Kubono}}}, \bibinfo {author} {\bibfnamefont
  {M.}~\bibnamefont {{Kunibu}}}, \bibinfo {author} {\bibfnamefont {Y.~U.}\
  \bibnamefont {{Matsuyama}}}, \bibinfo {author} {\bibfnamefont
  {S.}~\bibnamefont {{Michimasa}}}, \bibinfo {author} {\bibfnamefont
  {T.}~\bibnamefont {{Minemura}}}, \bibinfo {author} {\bibfnamefont
  {H.}~\bibnamefont {{Murakami}}}, \bibinfo {author} {\bibfnamefont
  {T.}~\bibnamefont {{Nakamura}}}, \bibinfo {author} {\bibfnamefont
  {A.}~\bibnamefont {{Saito}}}, \bibinfo {author} {\bibfnamefont
  {H.}~\bibnamefont {{Sakurai}}}, \bibinfo {author} {\bibfnamefont
  {M.}~\bibnamefont {{Serata}}}, \bibinfo {author} {\bibfnamefont
  {S.}~\bibnamefont {{Shimoura}}}, \bibinfo {author} {\bibfnamefont
  {T.}~\bibnamefont {{Sugimoto}}}, \bibinfo {author} {\bibfnamefont
  {E.}~\bibnamefont {{Takeshita}}}, \bibinfo {author} {\bibfnamefont
  {S.}~\bibnamefont {{Takeuchi}}}, \bibinfo {author} {\bibfnamefont
  {K.}~\bibnamefont {{Ue}}}, \bibinfo {author} {\bibfnamefont {K.}~\bibnamefont
  {{Yamada}}}, \bibinfo {author} {\bibfnamefont {Y.}~\bibnamefont
  {{Yanagisawa}}}, \bibinfo {author} {\bibfnamefont {A.}~\bibnamefont
  {{Yoshida}}}, \ and\ \bibinfo {author} {\bibfnamefont {M.}~\bibnamefont
  {{Ishihara}}},\ }\href@noop {} {\bibfield  {journal} {\bibinfo  {journal}
  {Prog.Theor.Phys.(Kyoto)}\ ,\ \bibinfo {pages} {Suppl. 146, 575}} (\bibinfo
  {year} {2002})}\BibitemShut {NoStop}%
\bibitem [{\citenamefont {Basunia}\ and\ \citenamefont
  {Hurst}(2016)}]{BASUNIA20161}%
  \BibitemOpen
  \bibfield  {author} {\bibinfo {author} {\bibfnamefont {M.}~\bibnamefont
  {Basunia}}\ and\ \bibinfo {author} {\bibfnamefont {A.}~\bibnamefont
  {Hurst}},\ }\href {\doibase https://doi.org/10.1016/j.nds.2016.04.001}
  {\bibfield  {journal} {\bibinfo  {journal} {Nuclear Data Sheets}\ }\textbf
  {\bibinfo {volume} {134}},\ \bibinfo {pages} {1} (\bibinfo {year}
  {2016})}\BibitemShut {NoStop}%
\bibitem [{\citenamefont {{Shamsuzzoha
  Basunia}}(2013)}]{SHAMSUZZOHABASUNIA20131189}%
  \BibitemOpen
  \bibfield  {author} {\bibinfo {author} {\bibfnamefont {M.}~\bibnamefont
  {{Shamsuzzoha Basunia}}},\ }\href {\doibase
  https://doi.org/10.1016/j.nds.2013.10.001} {\bibfield  {journal} {\bibinfo
  {journal} {Nuclear Data Sheets}\ }\textbf {\bibinfo {volume} {114}},\
  \bibinfo {pages} {1189} (\bibinfo {year} {2013})}\BibitemShut {NoStop}%
\bibitem [{\citenamefont {{Shamsuzzoha
  Basunia}}(2010)}]{SHAMSUZZOHABASUNIA20102331}%
  \BibitemOpen
  \bibfield  {author} {\bibinfo {author} {\bibfnamefont {M.}~\bibnamefont
  {{Shamsuzzoha Basunia}}},\ }\href {\doibase
  https://doi.org/10.1016/j.nds.2010.09.001} {\bibfield  {journal} {\bibinfo
  {journal} {Nuclear Data Sheets}\ }\textbf {\bibinfo {volume} {111}},\
  \bibinfo {pages} {2331} (\bibinfo {year} {2010})}\BibitemShut {NoStop}%
\bibitem [{\citenamefont {{Dufour}}\ \emph {et~al.}(1986)\citenamefont
  {{Dufour}}, \citenamefont {{Del Moral}}, \citenamefont {{Fleury}},
  \citenamefont {{Hubert}}, \citenamefont {{Jean}}, \citenamefont
  {{Pravikoff}}, \citenamefont {{Delagrange}}, \citenamefont {{Geissel}},\ and\
  \citenamefont {{Schmidt}}}]{NSR1986DU07}%
  \BibitemOpen
  \bibfield  {author} {\bibinfo {author} {\bibfnamefont {J.~P.}\ \bibnamefont
  {{Dufour}}}, \bibinfo {author} {\bibfnamefont {R.}~\bibnamefont {{Del
  Moral}}}, \bibinfo {author} {\bibfnamefont {A.}~\bibnamefont {{Fleury}}},
  \bibinfo {author} {\bibfnamefont {F.}~\bibnamefont {{Hubert}}}, \bibinfo
  {author} {\bibfnamefont {D.}~\bibnamefont {{Jean}}}, \bibinfo {author}
  {\bibfnamefont {M.~S.}\ \bibnamefont {{Pravikoff}}}, \bibinfo {author}
  {\bibfnamefont {H.}~\bibnamefont {{Delagrange}}}, \bibinfo {author}
  {\bibfnamefont {H.}~\bibnamefont {{Geissel}}}, \ and\ \bibinfo {author}
  {\bibfnamefont {K.~H.}\ \bibnamefont {{Schmidt}}},\ }\href@noop {} {\bibfield
   {journal} {\bibinfo  {journal} {Z.Phys.}\ }\textbf {\bibinfo {volume}
  {A324}},\ \bibinfo {pages} {487} (\bibinfo {year} {1986})}\BibitemShut
  {NoStop}%
\bibitem [{\citenamefont {Wu}(2000)}]{WU20001}%
  \BibitemOpen
  \bibfield  {author} {\bibinfo {author} {\bibfnamefont {S.-C.}\ \bibnamefont
  {Wu}},\ }\href {\doibase https://doi.org/10.1006/ndsh.2000.0014} {\bibfield
  {journal} {\bibinfo  {journal} {Nuclear Data Sheets}\ }\textbf {\bibinfo
  {volume} {91}},\ \bibinfo {pages} {1} (\bibinfo {year} {2000})}\BibitemShut
  {NoStop}%
\bibitem [{\citenamefont {Burrows}(2006)}]{BURROWS20061747}%
  \BibitemOpen
  \bibfield  {author} {\bibinfo {author} {\bibfnamefont {T.}~\bibnamefont
  {Burrows}},\ }\href {\doibase https://doi.org/10.1016/j.nds.2006.05.005}
  {\bibfield  {journal} {\bibinfo  {journal} {Nuclear Data Sheets}\ }\textbf
  {\bibinfo {volume} {107}},\ \bibinfo {pages} {1747} (\bibinfo {year}
  {2006})}\BibitemShut {NoStop}%
\bibitem [{\citenamefont {Chen}\ and\ \citenamefont {Singh}(2019)}]{CHEN20191}%
  \BibitemOpen
  \bibfield  {author} {\bibinfo {author} {\bibfnamefont {J.}~\bibnamefont
  {Chen}}\ and\ \bibinfo {author} {\bibfnamefont {B.}~\bibnamefont {Singh}},\
  }\href {\doibase https://doi.org/10.1016/j.nds.2019.04.001} {\bibfield
  {journal} {\bibinfo  {journal} {Nuclear Data Sheets}\ }\textbf {\bibinfo
  {volume} {157}},\ \bibinfo {pages} {1} (\bibinfo {year} {2019})}\BibitemShut
  {NoStop}%
\bibitem [{\citenamefont {Dong}\ and\ \citenamefont
  {Junde}(2015)}]{DONG2015185}%
  \BibitemOpen
  \bibfield  {author} {\bibinfo {author} {\bibfnamefont {Y.}~\bibnamefont
  {Dong}}\ and\ \bibinfo {author} {\bibfnamefont {H.}~\bibnamefont {Junde}},\
  }\href {\doibase https://doi.org/10.1016/j.nds.2015.08.003} {\bibfield
  {journal} {\bibinfo  {journal} {Nuclear Data Sheets}\ }\textbf {\bibinfo
  {volume} {128}},\ \bibinfo {pages} {185} (\bibinfo {year}
  {2015})}\BibitemShut {NoStop}%
\bibitem [{\citenamefont {Blachot}(2007)}]{BLACHOT20072035}%
  \BibitemOpen
  \bibfield  {author} {\bibinfo {author} {\bibfnamefont {J.}~\bibnamefont
  {Blachot}},\ }\href {\doibase https://doi.org/10.1016/j.nds.2007.09.001}
  {\bibfield  {journal} {\bibinfo  {journal} {Nuclear Data Sheets}\ }\textbf
  {\bibinfo {volume} {108}},\ \bibinfo {pages} {2035} (\bibinfo {year}
  {2007})}\BibitemShut {NoStop}%
\bibitem [{\citenamefont {{De Frenne}}\ and\ \citenamefont
  {Negret}(2008)}]{DEFRENNE2008943}%
  \BibitemOpen
  \bibfield  {author} {\bibinfo {author} {\bibfnamefont {D.}~\bibnamefont {{De
  Frenne}}}\ and\ \bibinfo {author} {\bibfnamefont {A.}~\bibnamefont
  {Negret}},\ }\href {\doibase https://doi.org/10.1016/j.nds.2008.03.002}
  {\bibfield  {journal} {\bibinfo  {journal} {Nuclear Data Sheets}\ }\textbf
  {\bibinfo {volume} {109}},\ \bibinfo {pages} {943} (\bibinfo {year}
  {2008})}\BibitemShut {NoStop}%
\bibitem [{\citenamefont {Silvestre-Brac}\ and\ \citenamefont
  {Piepenbring}(1982)}]{PhysRevC.26.2640}%
  \BibitemOpen
  \bibfield  {author} {\bibinfo {author} {\bibfnamefont {B.}~\bibnamefont
  {Silvestre-Brac}}\ and\ \bibinfo {author} {\bibfnamefont {R.}~\bibnamefont
  {Piepenbring}},\ }\href {\doibase 10.1103/PhysRevC.26.2640} {\bibfield
  {journal} {\bibinfo  {journal} {Phys. Rev. C}\ }\textbf {\bibinfo {volume}
  {26}},\ \bibinfo {pages} {2640} (\bibinfo {year} {1982})}\BibitemShut
  {NoStop}%
\end{thebibliography}%

\end{document}